\documentclass[11pt, onecolumn]{article}
\usepackage[top=1in, bottom=1in, left=1.25in, right=1.25in]{geometry}

\usepackage{amsfonts}
\usepackage{amsmath}
\usepackage{graphicx}
\usepackage{color, soul}
\usepackage{algorithm,algorithmic}
\usepackage{bm}
\usepackage{booktabs}
\usepackage{hyperref}
\usepackage{cite}

\usepackage[amsmath,thmmarks]{ntheorem}
\usepackage{theorem}

\usepackage{graphicx}

\newtheorem{pro}{Proposition}
\newtheorem{cor}{Corollary}

\newtheorem{defi}{Definition}
\newtheorem{rem}{Remark}
\newtheorem{thm}{Theorem}

\theoremheaderfont{\sc}\theorembodyfont{\upshape}
\theoremstyle{nonumberplain}
\theoremseparator{}
\theoremsymbol{\rule{1ex}{1ex}}

\renewcommand{\arraystretch}{1.5}

\newcommand{\figurewidth}{0.7\textwidth}

\begin{document}
\title{A Distributed Tracking Algorithm for Reconstruction of Graph Signals}

\author{Xiaohan~Wang\thanks{
Xiaohan Wang and Yuantao Gu are with the Department of Electronic Engineering, Tsinghua University, Beijing 100084, CHINA. The corresponding author of this paper is Yuantao Gu (gyt@tsinghua.edu.cn).},~Mengdi~Wang\thanks{
Mengdi Wang is with the Department of Operations Research and Financial Engineering, Princeton University, Princeton, NJ, 08544, USA.},~and~Yuantao~Gu$^*$}

\date{Received July 01, 2014; revised Nov. 24, 2014; accepted Jan. 31, 2015;\\
to appear in \emph{IEEE Journal of Selected Topics in Signal Processing}}

\maketitle

\begin{abstract}
The rapid development of signal processing on graphs provides a new perspective for processing large-scale data associated with irregular domains. In many practical applications, it is necessary to handle massive data sets through complex networks, in which most nodes have limited computing power. Designing efficient distributed algorithms is critical for this task. This paper focuses on the distributed reconstruction of a time-varying bandlimited graph signal based on observations sampled at a subset of selected nodes. A distributed least square reconstruction (DLSR) algorithm is proposed to recover the unknown signal iteratively, by allowing neighboring nodes to communicate with one another and make fast updates.  DLSR uses a decay scheme to annihilate the out-of-band energy occurring in the reconstruction process, which is inevitably caused by the transmission delay in distributed systems. Proof of convergence and error bounds for DLSR are provided in this paper, suggesting that the algorithm is able to track time-varying graph signals and perfectly reconstruct time-invariant signals. The DLSR algorithm is numerically experimented with synthetic data and real-world sensor network data, which verifies its ability in tracking slowly time-varying graph signals.

\textbf{Keywords}: Signal processing on graph, graph signal, distributed algorithm, sampling and reconstruction, time-varying signal.
\end{abstract}

\section{Introduction}

\subsection{Signal Processing on Graph}
During the past few years, the emerging field of signal processing on graphs (see \cite{shuman_emerging_2013,sandryhaila_discrete_2013}) has attracted vast research interests from multiple disciplines. Consider an $N$-vertex undirected graph, denoted as $\mathcal{G}(\mathcal{V}, \mathcal{E})$, where $\mathcal{V}$ is the set of vertices with $|\mathcal{V}|=N$ and $\mathcal{E}$ is
the set of edges.
A {\it graph signal} ${\bf f} \in \Re^N$ is a vector that assigns each vertex a real number.
Equivalently, the vector ${\bf f}$ is often regarded as a function $f: \mathcal{V}\mapsto \Re$.

Graph-based signal processing has been developed to analyze data or signal associated with irregular domains, e.g., large-scale networks. It finds wide applications in sensor networks \cite{zhu_graph_2012}, image processing \cite{narang_graph_2012}, recommendation systems \cite{narang_signal_2013}, etc.
Existing research topics on graph signal processing include:
graph signal sampling \cite{narang_localized_2013, anis_towards_2014},
uncertainty principle \cite{agaskar_aspectral_2013},
graph filtering \cite{chen_adaptive_2013},
spectral graph wavelet \cite{hammond_wavelets_2011, narang_perfect_2012},
graph signal compression \cite{zhu_approximating_2012},
graph signal multiresolution \cite{shuman_aframework_2013,ekambaram_multiresolution_2013},
parametric dictionary learning \cite{thanou_parametric_2013},
graph signal coarsening \cite{liu_coarsening_2014,liu_graphcoarsening_2014}, etc.

\subsection{Problem Description and Related Works}
In this work, we study the distributed reconstruction of smooth graph signals based on sample measurements obtained at representative nodes.
Suppose that the unknown graph signal lies in the low-frequency subspace, our reconstruction problem is to recover its missing entries from its known data/signal values sampled from a representative set of nodes.

Some theoretical results have been established for the sampling problem of bandlimited graph-based signals; see e.g., \cite{pesenson_sampling_2008,pesenson_variational_2009,pesenson_sampling_2010}.
The relation between the sample size necessary to obtain unique reconstruction and the cutoff frequency of bandlimited signal space has been studied.
Similar to classical results on time-domain irregular sampling, the idea of ``frame" has been introduced for graph signal processing.
The unique reconstruction conditions have been derived for normalized and unnormalized Laplacians \cite{pesenson_sampling_2008, pesenson_sampling_2010}.
As the field of graph signal processing is rapidly developing, we summarize some recent related works as follows.
A least square approach has been proposed in \cite{narang_signal_2013} to reconstruct bandlimited graph signal from signal values observed on sampled vertices, using a centralized algorithm.
An iterative method of bandlimited graph signal reconstruction has been proposed in \cite{narang_localized_2013}, with the practical consideration of balancing a tradeoff between smoothness and data-fitting.
Two more efficient iterative reconstruction methods using the local set have been considered in \cite{wang_iterative_2014,wang_localset_2014}.
A necessary and sufficient condition for perfect reconstruction of bandlimited graph signal has been derived in \cite{anis_towards_2014}.
Readers may refer to Section \ref{PreWork} for more details of related works.

Signal processing on graph is naturally related to distributed systems. For large-scale systems in lack of a central controller, e.g., sensor networks, distributed estimation and tracking \cite{bar_multi_1995} is an important topic. Algorithmic frameworks for distributed regression \cite{guestrin_distributed_2004} and inference \cite{paskin_arobust_2005} have been studied to fit global functions based on local measurements in sensor networks.
Consensus-based methods have been proposed in \cite{xiao_ascheme_2005,schizas_concensus_2008} to distributively compute the maximum likelihood estimate of unknown parameters.
Distributed Kalman filtering has been introduced in \cite{olfati_distributed_2007} for target tracking of sensor networks.
Diffusion RLS \cite{cattivelli_rls_2008} and LMS \cite{cattivelli_lms_2010} algorithms have been proposed for distributed estimation over adaptive networks.
To the best knowledge of the authors, there have been few works on the distributed reconstruction problem of bandlimited graph signal reconstruction.
A related work \cite{shuman_chebyshev_2011} proposes an approximation method that calculates the graph Fourier multipliers distributively, which we will discuss in subsequent sections.

In many practical distributed systems, a central processor is lacking and the majority of nodes have severely limited data processing power.
Moreover, the node-to-node transmission delay is non-negligible in large-scale networks, i.e., a given node cannot communicate globally with all other nodes to obtain instant fresh data.
These difficulties with large distributed systems pose a new and practical challenge to graph-based signal processing: how to reconstruct graph signals distributively and efficiently?
This is the motivation of the present paper, in which we propose a distributed algorithmic solution and answer the prior question to a reasonable extent.

\subsection{Contributions}

In this paper, we focus on the distributed recovery problem of graph-based time-varying signal.
We propose a distributed algorithm, namely the {\it distributed least square reconstruction} (DLSR), to adaptively reconstruct the missing values of a graph signal by allowing neighboring nodes to communicate with one another and make local updates.
Due to the transmission delay caused by node-to-node communication, some out-of-band energy inevitably occurs during the distributed signal reconstruction process.
The DLSR algorithm uses a decay factor to dampen this out-of-band energy and achieves {\it perfect reconstruction} of the unknown graph signal.
Theoretical convergence proof and error bounds of DLSR is given in our analysis.

The rest of this paper is organized as follows. In Section II, some preliminaries are introduced, including the basics of graph signal processing and a review of existing related works. In Section III, the distributed reconstruction algorithm (DLSR) is proposed and described in detail. In Section IV, the proof of convergence and error bound analysis for the proposed algorithm is presented. In Section V, DLSR is evaluated using numerical experiments with synthetic as well as real-world data.
\section{Preliminaries}
\subsection{Laplacian-Based Graph Signal Processing}
The concept of graph Laplacian is widely adopted in spectral graph theory \cite{chung_spectral_1997} and signal processing on graphs \cite{shuman_emerging_2013}.
For an undirected graph, the graph Laplacian (or unnormalized Laplacian, combinatorial Laplacian) is defined as
$$\bf{L=D-A},$$
where $\bf{A}$ is the adjacency matrix and $\bf{D}$ is the diagonal degree matrix, whose elements are the degrees of the corresponding vertices.
The normalized Laplacian is defined as
$${\bf \mathcal{L}=D}^{-\frac{1}{2}}{\bf LD}^{-\frac{1}{2}}.$$
Both unnormalized Laplacian and normalized Laplacian are real symmetric positive-semidefinite matrices and all the eigenvalues are nonnegative \cite{chung_spectral_1997}.
In the rest of the paper, we mainly focus on the normalized Laplacian. However, we note that analogous results can be easily obtained for unnormalized Laplacian.

In view of signal processing on graphs, the eigenvalues $\{\lambda_k\}$ of the Laplacian are regarded as frequencies and the corresponding eigenvectors $\{{\bf u}_k\}$ are regarded as basis vectors.
Consider an arbitrary graph signal ${\bf f}\in\Re^N$. Its frequency component corresponding to $\lambda_k$ is the inner product between ${\bf f}$ and the eigenvector ${\bf u}_k$, denoted as
$$
\hat{f}(\lambda_k)=\langle {\bf f}, {\bf u}_k\rangle=\sum_{i=1}^Nf(i)u_k(i).
$$

The eigenvectors associated with small eigenvalues have similar values on neighboring vertices, while the eigenvectors associated with large eigenvalues are the opposite.
As a result, the frequency components associated with small and large eigenvalues correspond to the low-frequency and high-frequency parts of the signal, respectively \cite{shuman_emerging_2013,biyikoglu_laplacian_2007}.

Suppose ${\mathbf f}\in\Re^N$ is a graph signal on an $N$-vertex graph $\mathcal{G}(\mathcal{V},\mathcal{E})$.
We say ${\mathbf f}$ is $\omega$-bandlimited if its frequency components corresponding to eigenvalues larger than $\omega$ are all zero.
In other words, the spectral support of ${\mathbf f}$ is a subset of $[0,\omega]$.
The subspace consisting of all $\omega$-bandlimited signals on graph $\mathcal{G}$ is called the Paley-Wiener space, which is a Hilbert space and denoted as $PW_{\omega}(\mathcal{G})$.

Suppose that for ${\mathbf f}\in PW_{\omega}(\mathcal{G})$ only the entries on a selected set of nodes $\{f(u)\}_{u\in \mathcal{S}}$ are known, where $\mathcal{S}\subseteq \mathcal{V}$ is the sampled vertex set.
The sampling and reconstruction problem is to recover the $\omega$-bandlimited original signal ${\bf f}$ based on the sampled data $\{f(u)\}_{u\in \mathcal{S}}$.

\subsection{Frame and Signal Reconstruction}
Bandlimited signal reconstruction is closely related to the frame theory. We briefly introduce its basics in the following.

\begin{defi}
A sequence of vectors $\{{\mathbf f}_i\}_{i\in \mathcal{I}}$ is a \emph{frame} in a Hilbert space $\mathcal{H}$, if there exist two constants $0<A\le B$ such that
$$
A\|{\mathbf f}\|^2\le\sum_{i\in \mathcal{I}}|\langle {\mathbf f}, {\mathbf f}_i\rangle|^2\le B\|{\mathbf f}\|^2,\quad \forall {\mathbf f}\in \mathcal{H}.
$$
Here the constants $A$ and $B$ are called frame bounds.
\end{defi}

\begin{defi}
For a frame $\{{\mathbf f}_i\}_{i\in \mathcal{I}}$, the \emph{frame operator}
${\bf S}: \mathcal{H}\rightarrow \mathcal{H}$
is
$$
{\mathbf{Sf}}=\sum_{i\in \mathcal{I}}\langle {\mathbf f}, {\mathbf f}_i\rangle {\mathbf f}_i,
$$
where ${\mathbf S}$ is invertible and satisfies $A{\mathbf I}\preceq {\mathbf S}\preceq B{\mathbf I}$.
\end{defi}

If the Euclidean matrix norm satisfies $\|{\bf I}-\lambda {\bf S}\|<1$, then
$$
{\mathbf f}={\bf S}^{-1}{\bf S}{\mathbf f}=\lambda \sum_{j=0}^{\infty}({\bf I}-\lambda {\bf S})^j{\bf S}{\mathbf f}.
$$
Consider the problem of reconstruction of an unknown signal ${\bf f}_*$. By defining
$$
{\mathbf f}^{(k)}=\lambda \sum_{j=0}^{k}({\bf I}-\lambda {\bf S})^j{\bf S}{\mathbf f}_*,
$$
we can use the following iteration to iteratively reconstruct ${\bf f}_*$ (see \cite{Christensen_an_2002}),
\begin{equation}\label{iteration}
{\mathbf f}^{(k+1)}=\lambda {\bf S}{\mathbf f}_*+({\bf I}-\lambda {\bf S}){\mathbf f}^{(k)}={\mathbf f}^{(k)}+\lambda {\bf S}({\mathbf f}_*-{\mathbf f}^{(k)}),
\end{equation}
which achieves an exponentially shrinking error bound
$$
\|{\mathbf f}^{(k)}-{\mathbf f}_*\|\le\|{\bf I}-\lambda {\bf S}\|^{k}\|{\mathbf f}^{(0)}-{\mathbf f}_*\|,\quad \forall k >0.
$$

\subsection{Previous Works on Bandlimited Graph Signal Reconstruction}\label{PreWork}
We review some basic concepts and important results regarding band limited graph signal reconstruction, which have been established in existing works.

\begin{defi}
\cite{pesenson_sampling_2008} A set of vertices $\mathcal{U}\subseteq \mathcal{V}(\mathcal{G})$ is a \emph{uniqueness set} for space $PW_{\omega}(\mathcal{G})$ if $\forall {\mathbf f}\in PW_{\omega}(\mathcal{G})$,
${\mathbf f}$ is uniquely determined by its values on $\mathcal{U}$, i.e., $\forall{\mathbf f},{\mathbf g}\in PW_{\omega}(\mathcal{G})$,
${\mathbf f}|_{\mathcal{U}}={\mathbf g}|_{\mathcal{U}}$ implies
${\mathbf f}={\mathbf g}$,
where ${\mathbf f}|_{\mathcal{U}}\in \Re^{|\mathcal{U}|}$ is the restriction of ${\bf f}$ to the subset $\mathcal{U}$.
\end{defi}

In order to perfectly reconstruct the bandlimited signals, we need the following relation between the sampling set and frame has been established.

\begin{thm}\label{thm3}
\cite{pesenson_sampling_2008} If the sampling set $\mathcal{S}$ is a uniqueness set for $PW_{\omega}(\mathcal{G})$, then $\{\mathcal{P}_{\omega}(\bm{\delta}_u)\}_{u\in \mathcal{S}}$ forms a frame in $PW_{\omega}(\mathcal{G})$, and the
upper bound $B=1$, where $\mathcal{P}_{\omega}(\cdot)$ is the orthogonal projection onto $PW_{\omega}(\mathcal{G})$, and $\bm{\delta}_u\in\Re^N$ is a graph signal on $\mathcal{G}$ whose entries satisfy
$$
\delta_u(v)=
\begin{cases}
1, & v=u; \\
0, & v\neq u.
\end{cases}
$$
\end{thm}

A method called iterative least square reconstruction (ILSR) has been proposed to reconstruct the bandlimited signal iteratively.
To be consistent with the results of our work, the method is rewritten as follows.

\begin{thm}
\cite{narang_localized_2013} If the sampling set $\mathcal{S}$ is a uniqueness set for $PW_{\omega}(\mathcal{G})$, then the original signal ${\bf f}_*\in PW_{\omega}(\mathcal{G})$ can be reconstructed using the sampled data $\{f_*(u)\}_{u\in\mathcal{S}}$
by the following ILSR method,
\begin{equation}\label{ILSR}
{\mathbf f}^{(k+1)}={\mathbf f}^{(k)}+\mathcal{P}_{\omega}\left(\sum_{u\in \mathcal{S}}(f_*(u)-f^{(k)}(u))\bm{\delta}_{u}\right),
\end{equation}
where ${\mathbf f}^{(k)}$ is a temporary result in the $k$th iteration.
\end{thm}

ILSR is derived from the method of projections onto convex sets (POCS) \cite{POCS1,POCS2}, which is also known as the alternating projection method.
The iteration (\ref{ILSR}) can be obtained by projecting onto the following two sets alternately,
\begin{align}
\mathcal{C}_1=&\{{\bf f}\in\Re^N|f(u)=f_*(u), \forall u\in\mathcal{S}\},\nonumber\\
\mathcal{C}_2=&PW_{\omega}(\mathcal{G}).\nonumber
\end{align}

An equivalent derivation of ILSR can be obtained with the help of frame theory.
Because of Theorem \ref{thm3}, the above method can also be obtained for $\lambda=1$ in the iteration (\ref{iteration}). Therefore, ${\mathbf f}^{(k)}$ in (\ref{ILSR}) is the same as that in (\ref{iteration}).

In addition to ILSR, two more efficient graph signal reconstruction methods, namely the iterative propagating reconstruction (IPR) and the iterative weighting reconstruction (IWR), have been proposed and proved to be convergent \cite{wang_iterative_2014,wang_localset_2014}.

Sampling and reconstruction of bandlimited graph signal is closely related to irregular sampling \cite{feichtinger_theory_1994,grochenig_adiscrete_1993,benedetto_irregular_1992,sauer_iterative_1987}
or non-uniform sampling \cite{marvasti_nonuniform_2001} in the time domain, which sheds light on the analysis of graph signal.
In fact, ILSR, IPR and IWR all have correspondence in time-domain irregular sampling, which are known as the Marvasti method \cite{marvasti_recovery_1991},
the Voronoi method \cite{grochenig_reconstruction_1992} and the adaptive weights method \cite{feichtinger_theory_1994}.

\subsection{Notation}
For a graph $\mathcal{G}$ and a cutoff frequency $\omega$, $PW_{\omega}(\mathcal{G})$ denotes the $\omega$-bandlimited space of graph signal on $\mathcal{G}$.
For any graph signal ${\bf f}\in\Re^N$, $\mathcal{P}_{\omega}({\bf f})$ denotes its orthogonal projection onto $PW_{\omega}(\mathcal{G})$, and $\mathcal{P}_{\omega+}({\bf f})$ denotes the projection onto the orthogonal complement space of $PW_{\omega}(\mathcal{G})$.
The sampling vertex set is denoted as $\mathcal{S}$, and the communication time delay between vertices $u$ and $v$ is denoted as $\tau (u,v)$.
It is assumed that one iteration is conducted at each time step.
The true graph signal at the $k$th time step or iteration is denoted as ${\bf f}_*^{(k)}$, whose entry associated with vertex $u$ is $f_*^{(k)}(u)$. $\tilde{\bf f}_*^{(k)}$ denotes a biased estimate of ${\bf f}_*^{(k)}$.
In the reconstruction algorithm, ${\bf f}^{(k)}$ denotes the temporary result obtained after the $k$th iteration, with $f^{(k)}(u)$ as its entry associated with vertex $u$.

\section{Distributed Reconstruction of Time-varying Bandlimited Graph Signal}

\subsection{Motivation}

We consider the distributed reconstruction of a time-varying low-frequency signal defined over graph by sampling at a small portion of nodes. The problem could be described in the scenario of wireless sensor network (WSN). For a given WSN, there are unknown function $f_*^{(t)}(v)$ associating with node $v$ at time $t$. Suppose that the function is slowly varying over both the time domain and the space domain. A snapshot of such function could be modeled as an unknown time-varying low-frequency signal ${\bf f}_*^{(t)}\in PW_{\omega}(\mathcal{G})$ located over a graph.\footnote{One may argue how one could model the WSN as a graph, e.g., how the weighted edges are yielded among all nodes. However, this problem is beyond the topic of this paper. We always assume the graph is available or could be estimated by some available methods.}

\begin{figure}
\centering
\includegraphics[width=0.55\textwidth]{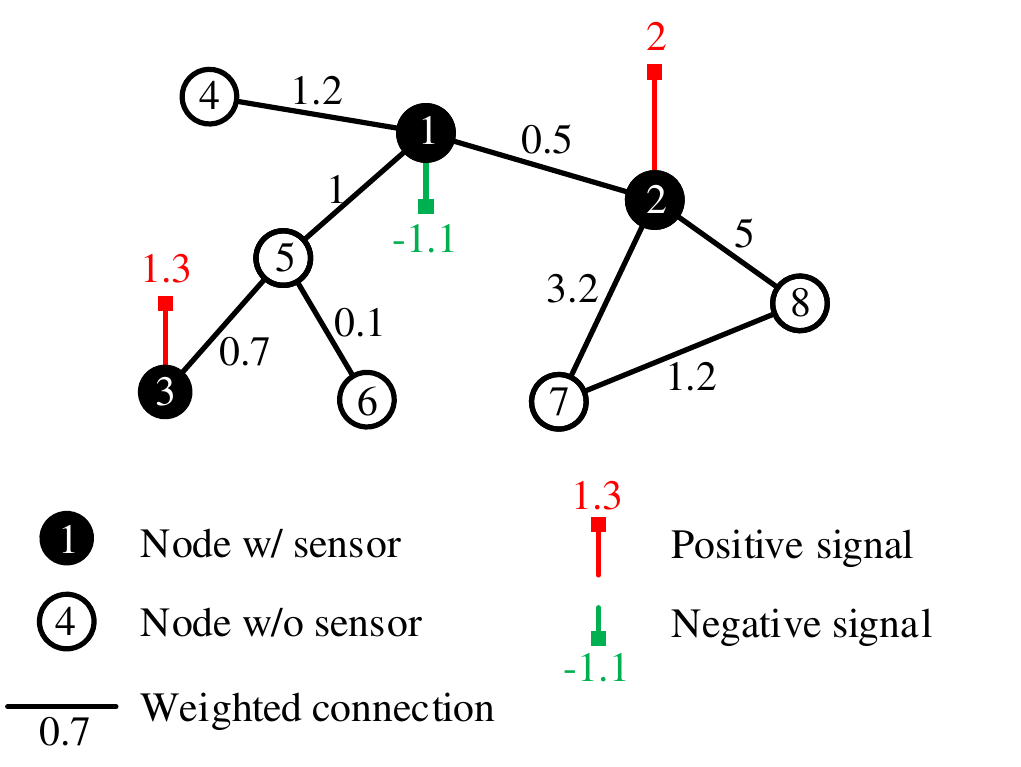}
\caption{An example of graph signal reconstruction over wireless sensor network.}
\label{Fig_systemdemo}
\end{figure}

Suppose that the WSN is a hybrid network and only a small subset of nodes in $\mathcal{S}$ are equipped with sensors. As a result, one can measure the signal entries on support $\mathcal{S}$ as $\{f_*^{(t)}(u)\}_{u\in\mathcal{S}}$ at time $t$. Our purpose is to distributivedly estimate the function values at all nodes $\{f_*^{(t)}(v)\}_{v\in\mathcal{V}}$, by using historical measurements at selected nodes $\{f_*^{(\tau)}(u)\}_{u\in\mathcal{S}, \tau\le t}$. See Fig. \ref{Fig_systemdemo} for a demonstration of the raised problem.

When the signal is time-invariant, the problem reduces to bandlimited graph signal reconstruction, where the nodes with and without sensors correspond to sampled and missing data.
In the distributed setting, the centralized iteration (\ref{ILSR}) no longer applies, as it is impossible for every node to obtain the instant estimation errors $\{f_*(u)-f^{(k)}(u)\}_{u\in\mathcal{S}}$.

In what follows, we focus on the generalization of ILSR method to distributed systems and time-varying signals. We proposed an algorithm called distributed least square reconstruction (DLSR). By letting each node conducting the iteration locally at each time instant, DLSR can adaptively reconstruct the missing entries of a slowly time-varying graph signal.

\subsection{Algorithm Description}

The basic idea of DLSR is to spread the current estimation errors associated with the representative nodes (which are equipped with sensors) to all other nodes over the connected network. Every node iteratively updates its own estimation based on its received messages.

The driver of the proposed algorithm is on those nodes with sensors, which calculates the error between the measurement $f_*^{(k)}(u)$ and the temporary estimation $f^{(k)}(u)$ in the $k$th iteration by
$$
	\epsilon^{(k)}(u) = f_*^{(k)}(u) - f^{(k)}(u), \quad\forall u\in\mathcal{S}.
$$
Then the estimation errors at node $u$ ($u\in\mathcal{S}$) are transmitted to other nodes in the network. At the $k$th iteration,  an arbitrary $v$ collects a set of delayed but most recent estimation errors,
$$
	\{\epsilon^{(k-\tau(u,v))}(u)\}_{u\in\mathcal{S}}, \quad \forall v\in\mathcal{V}(\mathcal{G}),
$$
where $\tau(u,v)$ denotes the transmission delay from node $u$ to node $v$
\footnote{For simplicity, we may set $e^{(k-\tau(u,v))}(u)=0$ if $k-\tau(u,v)<0$.}.
We denote the maximal transmission delay of the network by
$$
\tau=\max_{u\in\mathcal{S}, v\in\mathcal{V}(\mathcal{G})}\tau(u,v).
$$
Utilizing the most recent estimation errors, node $v$ updates its local estimate by
\begin{align}
f^{(k+1)}(v)=(1-\mu_{k+1}\beta_{k+1})f^{(k)}(v)
+\mu_{k+1}\sum_{u\in\mathcal{S}}\epsilon^{(k-\tau(u,v))}(u)(\mathcal{P}_{\omega}\bm{\delta}_u)(v),\quad \forall v\in\mathcal{V}(\mathcal{G})
\end{align}
where $\mu_{k+1}$ and $\beta_{k+1}$ denote the stepsize and decay factor, respectively. $(\mathcal{P}_{\omega}\bm{\delta}_u)(v)$ denotes the entry at $v$ of the lowpass component of $\bm{\delta}_u$, which could be calculated and stored before the system starts.

Please refer to Table \ref{Table_DLSR_u} and Table \ref{Table_DLSR_v}, which describe the detailed iterative process of signal reconstruction at the presentative nodes and the remaining nodes, respectively.

\begin{table}[t]
\renewcommand{\arraystretch}{1.2}
\caption{DLSR Algorithm at Representative Node $u\in\mathcal{S}$.}\label{Table_DLSR_u}
\begin{center}
\begin{tabular}{l}
\toprule[1pt]
{\bf Parameter:} \hspace{0.5em} $\mathcal{S}, \{\tau(u',u)\}_{u', u\in\mathcal{S}}, \mu_k, \beta_k$;\\
{\bf Initialization:} \hspace{0.5em} $f^{(0)}(u) = 0$, calculate $(\mathcal{P}_{\omega}\bm{\delta}_{u'})(u), \forall u'\in\mathcal{S}$;\\
\hline
{\bf For} $k=0,1,2,\cdots$\\
\hspace{1.3em} 1) Input: $f_*^{(k)}(u)$;\\
\hspace{1.3em} 2) Estimation:\\
\hspace{2.6em} $\epsilon^{(k)}(u) = f_*^{(k)}(u) - f^{(k)}(u)$;\\
\hspace{1.3em} 3) Communication:\\
\hspace{2.6em} Send $\epsilon^{(k)}(u)$ and $\epsilon^{(k-1-\tau(u',u))}(u')$ to neighbors, $\forall u'\in\mathcal{S}\backslash u$;\\
\hspace{2.6em} Receive $\epsilon^{(k-\tau(u',u))}(u')$ from neighbors, $\forall u'\in\mathcal{S}\backslash u$;\\
\hspace{1.3em} 4) Update Storage:\\
\hspace{2.6em} Save $\epsilon^{(k-\tau(u',u))}(u')$, $\forall u'\in\mathcal{S}\backslash u$;\\
\hspace{1.3em} 5) Update Estimation:\\
\hspace{2.6em} $f^{(k+1)}(u)=(1-\mu_{k+1}\beta_{k+1})f^{(k)}(u)$\\
\hspace{9em}$+ \mu_{k+1}\sum\limits_{u'\in\mathcal{S}}\epsilon^{(k-\tau(u',u))}(u')(\mathcal{P}_{\omega}\bm{\delta}_{u'})(u)$.\\
{\bf End}\\
\bottomrule[1pt]
\end{tabular}
\end{center}
\end{table}

\begin{table}[t]
\renewcommand{\arraystretch}{1.2}
\caption{DLSR Algorithm at Non-representative Node $v\in\mathcal{V}(\mathcal{G})\backslash\mathcal{S}$.}\label{Table_DLSR_v}
\begin{center}
\begin{tabular}{l}
\toprule[1pt]
{\bf Parameter:} \hspace{0.5em} $\mathcal{S}, \{\tau(u,v)\}_{u\in\mathcal{S}, v\in\mathcal{V}(\mathcal{G})\backslash\mathcal{S}}, \mu_k, \beta_k$;\\
{\bf Initialization:} \hspace{0.5em} $f^{(0)}(v) = 0$, calculate $(\mathcal{P}_{\omega}\bm{\delta}_{u})(v), \forall u\in\mathcal{S}$;\\
\hline
{\bf For} $k=0,1,2,\cdots$\\
\hspace{1.3em} 1) Communication:\\
\hspace{2.6em} Send $\epsilon^{(k-1-\tau(u,v))}(u)$ to neighbors, $\forall u\in\mathcal{S}$;\\
\hspace{2.6em} Receive $\epsilon^{(k-\tau(u,v))}(u)$ from neighbors, $\forall u\in\mathcal{S}$;\\
\hspace{1.3em} 2) Update Storage:\\
\hspace{2.6em} Save $\epsilon^{(k-\tau(u,v))}(u)$, $\forall u\in\mathcal{S}$;\\
\hspace{1.3em} 3) Update Estimation:\\
\hspace{2.6em} $f^{(k+1)}(v)=(1-\mu_{k+1}\beta_{k+1})f^{(k)}(v)$\\
\hspace{9em}$+\mu_{k+1}\sum\limits_{u\in\mathcal{S}}\epsilon^{(k-\tau(u,v))}(u)(\mathcal{P}_{\omega}\bm{\delta}_{u})(v)$;\\
\hspace{1.3em} 4) Output: $f^{(k+1)}(v)$.\\
{\bf End}\\
\bottomrule[1pt]
\end{tabular}
\end{center}
\end{table}

\subsection{An Example}

In order to provide some intuition for our distributed algorithm,  let us refer to Fig. \ref{Fig_systemdemo} and describe what happens on a typical node in the network.

\begin{itemize}
\item
As a representative node equipped with sensor,  node $1$ will get a measurement $f_*^{(k)}(1)$ at the $k$th iteration and then calculate the estimation error $\epsilon^{(k)}(1)$. The estimation error will be send to its neighbors of node $2, 4$, and $5$, and then forwarded to others. At the same slot, node $1$ will receive the estimation errors of other nodes with sensors and use the most recent ones ($\epsilon^{(k-1)}(2)$ from node $2$ and $\epsilon^{(k-2)}(3)$ from node $5$). Consequently, it could update the estimation by
\begin{align*}
	f^{(k+1)}(1) &= (1-\mu_{k+1}\beta_{k+1})f^{(k)}(1) \\
	&\quad+ \mu_{k+1}\Big( \epsilon^{(k)}(1)\cdot(\mathcal{P}_{\omega}\bm{\delta}_{1})(1)
	 + \epsilon^{(k-1)}(2)(\mathcal{P}_{\omega}\bm{\delta}_{2})(1) + \epsilon^{(k-2)}(3)(\mathcal{P}_{\omega}\bm{\delta}_{3})(1)\Big).
\end{align*}
One may notice that the new estimation is not the output of the proposed algorithm, because our purpose is to estimate the strength of the signal associated with the node without sensor. However, the new estimation errors at representative nodes will be transmitted over the network to help all others to conduct their estimation.
\item
As a regular node that is not equipped with sensor, at the $k$th iteration, node $5$ will receive $\epsilon^{(k-1)}(1), \epsilon^{(k-2)}(2)$, and $\epsilon^{(k-1)}(3)$ from its neighbors. Then it will transmit the most recent estimation errors of nodes with sensors to its neighbors $1, 3,$ and $6$.
The estimate of node $5$ is updated by
\begin{align*}
	f^{(k+1)}(5) &= (1-\mu_{k+1}\beta_{k+1})f^{(k)}(5) \\
	&\quad + \mu_{k+1}\Big( \epsilon^{(k-1)}(1)(\mathcal{P}_{\omega}\bm{\delta}_{1})(5)
	+ \epsilon^{(k-2)}(2)(\mathcal{P}_{\omega}\bm{\delta}_{2})(5)
	+ \epsilon^{(k-3)}(3)(\mathcal{P}_{\omega}\bm{\delta}_{3})(5)\Big).
\end{align*}
The new estimate will be sent out as a temporary result of the proposed algorithm.
\end{itemize}

\subsection{Discussions}

Asynchronization is one of the most common issues in distributed systems. In our distributed reconstruction setting, asynchronization leads to communication delay between nodes that are connected through multiple links, which induces a deviation of the estimated signal from the bandlimited space. However, as long as the maximum delay in the network is bounded by a constant $\tau$, the proposed method can successfully annihilate the out-of-band estimation error and achieves perfect reconstruction.

Node failure is also a common problem in WSN. The proposed DLSR is robust to both communication failure and sensor failure.
In the former case, some links are broken and fail to work. The data packets have to be delivered through new route and the transmission delay may increase. Even in this case, the DLSR is still going to work, provided that the network remains connected and the maximum transmission delay is bounded. In the case of a sensor failure, some presentative nodes (i. e. $u\in\mathcal{S}$) no longer obtain the sampled data, which means that they can act the same as the regular nodes. As long as the system is designed with some redundancy, the DLSR can still work, provided that there remain enough number of functional sensors.

The proposed algorithm requires that the vectors $\{\mathcal{P}_{\omega}\bm{\delta}_u\}_{u\in\mathcal{S}}$ be calculated in advance and their entries be stored in the respective nodes. In some practical situation this pre-calculation could be unavailable. However, a distributed method proposed by \cite{shuman_chebyshev_2011} can be used to calculate $\{(\mathcal{P}_{\omega}\bm{\delta}_u)(v)\}_{u\in\mathcal{S}}$ approximately at node $v$.

In fact, the operation $\mathcal{P}_{\omega}(\cdot)$ for any given graph signal can be approximately calculated by a distributed method proposed by \cite{shuman_chebyshev_2011}. By this method, it will take some (depends on the network scale) rounds of data transmission and calculation to obtain the approximate projection. Therefore, another distributed method can be readily proposed by directly applying the above approximate projection to ILSR with a stepsize $\mu$ to track time-varying signal.
Our method differs from this approach in the following aspects.
\begin{itemize}
\item
Supposing the period of a time step composed of one transmission and calculation is fixed in both methods,
it will take some time steps to implement one iteration of ILSR by directly applying the method in \cite{shuman_chebyshev_2011} to ILSR. Therefore, the data used in the calculation are all sampled several time steps earlier.
In our method, the iteration is conducted in one time step and use the current data at every node, which means the samples are as fresh as possible, the delay is only caused by transmission, and there is no waiting. Although nonuniform delays may violate the bandlimited property, we will prove in the following section that the out-of-band energy can be eliminated eventually.
\item
The projection $\mathcal{P}_{\omega}(\cdot)$ should be conducted for different graph signals as the iteration goes on by directly applying the method in \cite{shuman_chebyshev_2011} to ILSR, and each projection takes some time steps.
In the proposed method, only pre-calculating (precisely or approximately using the method in \cite{shuman_chebyshev_2011}) the frame elements, $\{\mathcal{P}_{\omega}\bm{\delta}_u\}_{u\in\mathcal{S}}$, are enough to reconstruct the graph signal, which is more economical.
\end{itemize}

\section{Convergence Analysis}

We will first study the convergence behavior of DLSR in a general situation that the bandlimited graph signal to be constructed varies slowly by time, and then specialize the result to a time-invariant case. In order to simplify the expression, we will fix stepsizes $\mu$ and $\beta$ to be constants in studying the time-varying case. Finally, we let the stepsizes be diminishing and show that DLSR achieves a perfect reconstruction of time-invariant signal.

\subsection{Tracking Time-Varying Signal Using Constant Parameters}

The proposed DLSR algorithm is equivalent to the following iteration in the vector form
\begin{equation}\label{DistIter}
{\bf f}^{(k+1)}=(1-\mu\beta){\bf f}^{(k)}+\mu\sum_{u\in\mathcal{S}}\left({\bf F}_{*u}^{(k)}-{\bf F}_u^{(k)}\right)\mathcal{P}_{\omega}\bm{\delta}_u,
\end{equation}
where
$$
{\bf F}_{*u}^{(k)}-{\bf F}_u^{(k)}=\text{diag}\left\{f_*^{(k-\tau(u,i))}(u)-f^{(k-\tau(u,i))}(u)\right\}_{i=1,\cdots,N}
$$
is a diagonal matrix composed of the delayed estimation error at node $u$.

Although $\mathcal{P}_{\omega}\bm{\delta}_u$ is bandlimited for any $u$,
by introducing ${\bf F}_u^{(k)}$, the delayed signal $({\bf F}_{*u}^{(k)}-{\bf F}_u^{(k)})\mathcal{P}_{\omega}\bm{\delta}_u$ no longer belongs to the low-frequency subspace $PW_{\omega}(\mathcal{G})$. As a result, the sequence of estimated signals $\{{\bf f}^{(k)}\}$ are no longer $\omega$-bandlimited. This existence of out-of-band energy makes DLSR critically different from its centralized version (\ref{ILSR}) and substantially complicates the convergence analysis. Since ${\bf f}^{(k)}$ is not $\omega$-bandlimited, we need to study its low-frequency and high-frequency components separately.

For given node set $\mathcal{S}$ and cutoff frequency $\omega$, we may define an operator ${\bf T}$ on a graph signal ${\bf f}$ as
\begin{align}\label{DefS}
{\bf Tf}&=\mathcal{P}_{\omega}\left(\sum_{u\in\mathcal{S}}f(u)\bm{\delta}_u\right)\\
&=\sum_{u\in\mathcal{S}}f(u)\mathcal{P}_{\omega}\bm{\delta}_u.\nonumber
\end{align}
According to Theorem \ref{thm3}, if $\mathcal{S}$ is the uniqueness set of graph $\mathcal{G}$ with respect to $\omega$, $\{\mathcal{P}_{\omega}\bm{\delta}_u\}_{u\in\mathcal{S}}$ is a frame in $PW_{\omega}(\mathcal{G})$.
For any ${\bf f}\in PW_{\omega}(\mathcal{G})$, using the fact that
$$
	f(u)=\langle \mathcal{P}_{\omega}{\bf f},\bm{\delta}_u\rangle=\langle {\bf f}, \mathcal{P}_{\omega}\bm{\delta}_u\rangle,\quad \forall u\in\mathcal{S},
$$
${\bf Tf}$ can be rewritten as
$$
{\bf Tf}=\sum_{u\in\mathcal{S}}\langle {\bf f}, \mathcal{P}_{\omega}\bm{\delta}_u\rangle\mathcal{P}_{\omega}\bm{\delta}_u,\quad
\forall {\bf f}\in PW_{\omega}(\mathcal{G}),
$$
which is the frame operator of $\{\mathcal{P}_{\omega}\bm{\delta}_u\}_{u\in\mathcal{S}}$, and the frame bounds are $A$ and $B$.

The definition of ${\bf T}$ implies
$$
\|{\bf Tf}\|\le\left\|\sum_{u\in\mathcal{S}}f(u)\bm{\delta}_u\right\|\le\|{\bf f}\|
$$
and one has $\|{\bf T}\|\le 1$. By defining $\tilde{\bf f}_*^{(k)}$ as
\begin{equation}\label{deftildefstar}
\tilde{\bf f}_*^{(k)}=(\beta {\bf I}+{\bf T})^{-1}{\bf Tf}_*^{(k)},
\end{equation}
one further gets
\begin{equation}\label{FStar}
{\bf Tf}_*^{(k)}=\beta \tilde{\bf f}_*^{(k)}+{\bf T}\tilde{\bf f}_*^{(k)}.
\end{equation}
According to (\ref{DefS}), both ${\bf Tf}_*^{(k)}$ and ${\bf T}\tilde{\bf f}_*^{(k)}$ are within the low-frequency space $PW_{\omega}(\mathcal{G})$. Therefore one can obtain from (\ref{FStar}) that $\tilde{\bf f}_*^{(k)}\in PW_{\omega}(\mathcal{G})$. As a consequence, $\mathcal{P}_{\omega_+}\tilde{\bf f}_*^{(k)}={\bf 0}$, where $\mathcal{P}_{\omega_+}$ denotes the projection operator onto the high-frequency subspace which is the orthogonal complement of $PW_{\omega}(\mathcal{G})$.

By defining the in-band error and out-of-band error as, respectively,
\begin{align}
e^{(k)}&=\left\|\mathcal{P}_{\omega}{\bf f}^{(k)}-\mathcal{P}_{\omega}\tilde{\bf f}_*^{(k)}\right\|,\label{eq:eofk}\\
e_+^{(k)}&=\left\|\mathcal{P}_{{\omega}_+}{\bf f}^{(k)}-\mathcal{P}_{{\omega}_+}\tilde{\bf f}_*^{(k)}\right\| =\left\|\mathcal{P}_{{\omega}_+}{\bf f}^{(k)}\right\|,\label{eq:epofk}
\end{align}
the following proposition gives inequalities that $\left\{e^{(k)}\right\}$ and $\{e_+^{(k)}\}$ satisfy.
Further, it will be shown that if the signal varies slowly enough, by properly selecting the stepsize $\mu$,
DLSR can track time-varying signals.

\begin{pro}\label{TimeVariant}
Supposing the true signal satisfies
\begin{equation}\label{kappa}
\left|f_*^{(k+1)}(u)-f_*^{(k)}(u)\right|\le\Delta, \quad\forall u\in\mathcal{V(G)}, k\ge1,
\end{equation}
if
$\Delta\le\min\left\{\Delta_{\text{max}},\beta B_{e_+}/(|\mathcal{S}|\tau)\right\}$
and
\begin{equation}\label{eq:boundsofmu}
\mu_{\text{min}}\le\mu<\min{\left\{\mu_{\text{max}},\frac{1}{\beta+A},\frac{B_{\eta}}{C+|\mathcal{S}|^{\frac{3}{2}}\tau^2\Delta}\right\}},
\end{equation}
the errors $\left\{e^{(k)}\right\}$ and $\{e_+^{(k)}\}$ satisfy
\begin{align}
e_+^{(k+1)} & \le (1-\mu\beta)e_+^{(k)}+\mu^2C+M(\mu)\Delta,\label{TVOutBandErr}\\
e^{(k+1)} &\le (1-\mu\beta-\mu A)e^{(k)}+\mu\|{\bf T}\|e_+^{(k)}+\mu^2C+M(\mu)\Delta.\label{TVInBandErr}
\end{align}
In the above inequalities,
\begin{equation}\label{eq:Mofmu}
M(\mu)=|\mathcal{S}|^{\frac{3}{2}}\tau^2\mu^2+|\mathcal{S}|\tau\mu+\sqrt{N},
\end{equation}
$\Delta_{\text{max}}$ is the positive root of
\begin{align}
|\mathcal{S}|^{\frac{3}{2}}\tau^2\left(4N^{\frac{1}{2}}-|\mathcal{S}|^{\frac{1}{2}}\right)\Delta^2
+\left(2|\mathcal{S}|\tau\beta B_{e_+} +4N^{\frac{1}{2}} C\right)\Delta
-\beta^2 B_{e_+}^2=0,\nonumber
\end{align}
$\mu_{\text{min}}$ and $\mu_{\text{max}}$ are the roots of
$$
\left(C+|\mathcal{S}|^{\frac{3}{2}}\tau^2\Delta\right)\mu^2+\left(|\mathcal{S}|\tau\Delta-\beta B_{e_+}\right)\mu+\sqrt{N}\Delta=0,
$$
$A$ and $B$ are the frame bounds of ${\bf T}$ in $PW_{\omega}(\mathcal{G})$, $\|{\bf T}\|$ is the norm of ${\bf T}$, $B_{\eta}$, $B_e$ and $B_{e_+}$ are constants satisfying
\begin{equation}\label{eq:additionalcondition}
(\beta+A)B_e=(\beta+\|{\bf T}\|)B_{e_+},
\end{equation}
and $C$ is a constant
\begin{equation}\label{DefC}
C=\tau\sqrt{|\mathcal{S}|}\left((\beta+\|{\bf T}\|)(B_e+B_{e_+})+B_{\eta}\right).
\end{equation}
\end{pro}

The proof of Proposition \ref{TimeVariant} is postponed to \ref{Proof}.

\begin{rem}
According to Proposition \ref{TimeVariant}, the out-of-band error (\ref{TVOutBandErr}) shows the necessity of the decay factor $\beta$.
If $\beta=0$, the out-of-band error cannot be proved to converge, and the error may accumulate with the iterations.
The decay factor $\beta$ enhances the robustness of iteration.
\end{rem}

\begin{rem}\label{rem2}
The inequality (\ref{TVInBandErr}) shows that the out-of-band error $e_+^{(k)}$ also affects the in-band error $e^{(k+1)}$, which implies that the in-band error cannot be very small if the out-of-band error exists. In other words, it is important to eliminate the out-of-band error.
\end{rem}

Taking the limit superior of (\ref{TVOutBandErr}) and (\ref{TVInBandErr}), one obtains
\begin{align}
\limsup_{k\rightarrow\infty}e_+^{(k)}&\le\left(D+\frac{E}{\beta}\right)\mu
+\frac{M(\mu)}{\beta\mu}\Delta,\label{TVLimsupEplus}\\
\limsup_{k\rightarrow\infty}e^{(k)}&\le\left(1+\frac{\|{\bf T}\|}{\beta}\right) \left(\frac{D\beta+E}{\beta+A}\mu+\frac{M(\mu)}{(\beta+A)\mu}\Delta\right).\label{TVLimsupE}
\end{align}
where $D$ and $E$ are constants. The above results imply that for a constant stepsize $\mu$, the out-of-band error $e_+^{(k)}$ will eventually get below a threshold that is determined by $\mu$ and $\Delta$. Similarly, $\mathcal{P}_{\omega}{\bf f}^{(k)}$ will converge to the neighborhood of $\mathcal{P}_{\omega}\tilde{\bf f}_*^{(k)}$, and the error is also controlled by $\mu$ and $\Delta$.

\begin{rem}\label{remnew}
Because bias is introduced by the multiple $1-\mu\beta$ in the iteration,
the low-frequency and high-frequency components of the temporary estimate will get into the neighborhoods of $\mathcal{P}_{\omega}\tilde{\bf f}_*^{(k)}$ and ${\bf 0}$, respectively. Because of the influence of the decay factor $\beta$, the reconstructed signal is biased. It will be proved in Corollary \ref{TimeInvariant} that in the time-invariant case, these two components will exactly converge to $\mathcal{P}_{\omega}\tilde{\bf f}_*^{(k)}$ and ${\bf 0}$.
\end{rem}

\subsection{Recovering Time-invariant Signal Using Constant Parameters}

For the time-invariant case, ${\bf f}_*^{(k)}$ can be written as ${\bf f}_*$ and ${\bf F}_{*u}^{(k)}$ becomes $f_*(u){\bf I}_N$.
Similar to (\ref{deftildefstar}), $\tilde{\bf f}_*$ can also be defined as
\begin{equation}\label{defTItildefstar}
\tilde{\bf f}_*=(\beta {\bf I}+{\bf T})^{-1}{\bf Tf}_*.
\end{equation}
Then Proposition \ref{TimeVariant} becomes the following corollary for $\Delta=0$.
\begin{cor}\label{TimeInvariant}
For time-invariant true signal ${\bf f}_*$, supposing the in-band error and out-of-band error are defined as, respectively,
\begin{align*}
e^{(k)}&=\|\mathcal{P}_{\omega}{\bf f}^{(k)}-\mathcal{P}_{\omega}\tilde{\bf f}_*\|,\\
e_+^{(k)}&=\|\mathcal{P}_{{\omega}_+}{\bf f}^{(k)}-\mathcal{P}_{{\omega}_+}\tilde{\bf f}_*\|=\|\mathcal{P}_{{\omega}_+}{\bf f}^{(k)}\|,
\end{align*}
the error $\{e^{(k)}\}$ and $\{e_+^{(k)}\}$ satisfy
\begin{align}
e_+^{(k+1)}&\le (1-\mu\beta)e_+^{(k)}+\mu^2C,\label{TIOutBandErr}\\
e^{(k+1)}&\le (1-\mu\beta-\mu A)e^{(k)}+\mu\|{\bf T}\|e_+^{(k)}+\mu^2C,\label{TIInBandErr}
\end{align}
if
$$
\mu<\min{\left\{\frac{1}{\beta+A},\frac{B_{\eta}}{C},\frac{\beta B_{e_+}}{C}\right\}},
$$
where the constants are the same as those in Proposition \ref{TimeVariant}.
\end{cor}

Besides, (\ref{TVLimsupEplus}) and (\ref{TVLimsupE}) become, respectively,
\begin{align}
\limsup_{k\rightarrow\infty}e_+^{(k)}&\le\frac{C}{\beta}\mu=\left(D+\frac{E}{\beta}\right)\mu,\label{TILimsupEplus}\\
\limsup_{k\rightarrow\infty}e^{(k)}&\le \frac{D\beta+E}{\beta+A}\left(1+\frac{\|{\bf T}\|}{\beta}\right)\mu.\label{TILimsupE}
\end{align}

\begin{rem}\label{remnew2}
For the time-invariant case, the out-of-band error $e_+^{(k)}$ will get below a threshold that is proportional to the stepsize $\mu$. It means that the out-of-band energy will be almost eliminated eventually along with the iteration if $\mu$ is small. $\mathcal{P}_{\omega}{\bf f}^{(k)}$ will converge to the neighborhood of $\mathcal{P}_{\omega}\tilde{\bf f}_*$, and its radius is also proportional to $\mu$. Therefore, for diminishing stepsize $\mu_k$ approaching $0$, $\mathcal{P}_{\omega}{\bf f}^{(k)}$ and $\mathcal{P}_{\omega_+}{\bf f}^{(k)}$ will strictly converge to $\mathcal{P}_{\omega}\tilde{\bf f}_*$ and ${\bf 0}$, respectively, for a sequence of properly chosen diminishing stepsize.
\end{rem}

The bias will be estimated next. Because ${\bf f}_*, \tilde{\bf f}_*\in PW_{\omega}(\mathcal{G})$, the iteration ${\bf f}^{(k)}$ will converge to $\tilde{\bf f}_*$.
Considering the definition of $\tilde{\bf f}_*$ in (\ref{defTItildefstar}), the bias satisfies
\begin{align}
\tilde{\bf f}_*-{\bf f}_*&=(\beta {\bf I}+{\bf T})^{-1}{\bf Tf}_*-{\bf f}_*\nonumber\\
&=(\beta {\bf I}+{\bf T})^{-1}{\bf Tf}_*-(\beta {\bf I}+{\bf T})^{-1}(\beta {\bf I}+{\bf T}){\bf f}_*\nonumber\\
&=-\beta(\beta {\bf I}+{\bf T})^{-1}{\bf f}_*.\nonumber
\end{align}
For ${\bf f}_*\in PW_{\omega}(\mathcal{G})$, according to the frame bounds of operator ${\bf T}$, we have
$$
\|(\beta {\bf I}+{\bf T})^{-1}{\bf f}_*\|\le\frac{1}{\beta+A}\|{\bf f}_*\|,
$$
and then
\begin{equation}\label{Bias}
\|\tilde{\bf f}_*-{\bf f}_*\|\le\frac{\beta}{\beta+A}\|{\bf f}_*\|.
\end{equation}
Thus, the bias is determined by $\beta$ and decreases with the decrease of $\beta$.

Combining (\ref{TILimsupEplus}), (\ref{TILimsupE}), and (\ref{Bias}), the following proposition gives the limit superior of the total error, as a function of $\beta$ and $\mu$.

\begin{pro}\label{pro2}
The total error for the time-invariant case satisfies
$$
\limsup_{k\rightarrow\infty}\|{\bf f}^{(k)}-{\bf f}_*\|\le F\beta+G\frac{\mu}{\beta}+H\mu+J\beta\mu,
$$
where $F$, $G$, $H$, and $J$ are constants.
\end{pro}

Proposition \ref{pro2} can be easily proved by summing up (\ref{TILimsupEplus}), (\ref{TILimsupE}), and (\ref{Bias}).

\subsection{Recovering Time-Invariant Signal Using Variable Parameters}

Finally we will back to the general situation of variable stepsize and decay factor. According to Proposition \ref{pro2}, by discarding the higher order of a diminishing stepsize $\mu_k$, the best $\beta_k$ satisfies $\beta_k\sim O(\sqrt{\mu_k})$ to minimize the total error. Accordingly, the total error bound can be controlled by adjusting $\mu_k$ and $\beta_k$. In other words, the reconstruction error can be made {\it arbitrarily small}: DRSL achieves perfect reconstruction.

The following proposition gives the convergence analysis for a special choice of diminishing stepsize and decay factor.

\begin{pro}\label{pro3}
For diminishing stepsize $\mu_k=\mu_1/\sqrt{k}$ and decay factor $\beta_k=\beta_1/\sqrt[4]{k}$, the total error satisfies the following inequality,
$$
\|{\bf f}^{(k)}-{\bf f}_*\|\le K/\sqrt[4]{k},
$$
where $K$ is a constant.
It means that the total estimation error decreases on the rate of $1/\sqrt[4]{k}$ and converges to zero eventually.
\end{pro}

The proof of Proposition \ref{pro3} is postponed to \ref{proof2}.

\section{Experiments}
Experiments are designed to confirm the theoretical analysis and test the performance of the proposed distributed algorithm.
The graph is generated by $100$ randomly located nodes and the edges are generated by $4$-nearest neighbors of the nodes,
and the weights are inversely proportional to the square of geometric distance.
Among the $100$ nodes, $20$ of them are randomly selected as the sampling set.
The cutoff frequency is chosen to guarantee that the sampling node set is a uniqueness set,
which can be determined by the method given in \cite{narang_signal_2013}.
\footnote{
Proposition 2 of \cite{narang_signal_2013}: The sampling set $\mathcal{S}$ is a unique set for $PW_{\omega}(\mathcal{G})$ if the cutoff frequency satisfies $\omega\le\sigma_{\text{min}}$, where $\sigma_{\text{min}}^2$ is the smallest singular value of
$({\mathcal{L}}^2)_{\mathcal{S}^{\text{c}}}$, which is the submatrix of $\mathcal{L}^2$ containing only the rows and columns corresponding to the complementary set of $\mathcal{S}$, and $\mathcal{L}$ is the normalized Laplacian of $\mathcal{G}$.
}
The bandlimited signal is generated by filtering the high-frequency components off.
The transmission delay of each pair of nodes is simply regarded as the number of hops between them in the graph. The maximal transmission delay of this graph is $14$.

\subsection{Tracking Time-Varying Signals}

\subsubsection{Tracking Performance}
\begin{figure}[t]
\begin{center}
\includegraphics[width=\figurewidth]{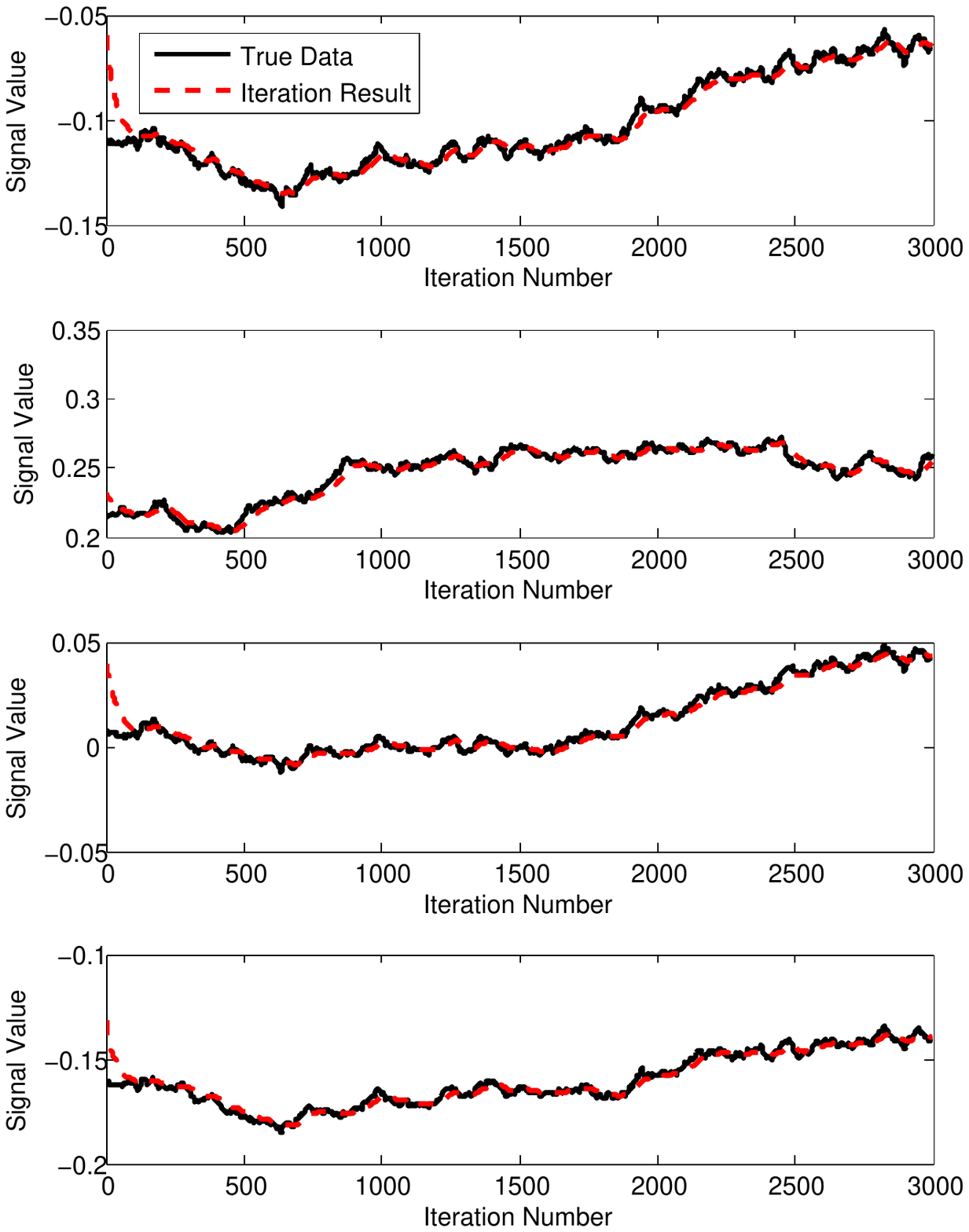}
\caption{Time-varying aiming signal and iterative results of four nodes. The proposed algorithm can track the aiming signal over time.}
\label{exp3}
\end{center}
\end{figure}

In this experiment, the tracking performance of DLSR is verified.
The parameters are chosen as $\Delta=0.005, \mu=0.1$, and $\beta=10^{-3}$.
The time-varying signal is generated by adding a random bandlimited increment whose largest absolute entry is $\Delta$ for each time step.
The aiming signal and iterative results of four nodes are focused on, as illustrated in Fig. \ref{exp3}.
The nodes associated with the upper two subfigures are in the sampling set, and the nodes in the lower two subfigures are not in the sampling set.
All the nodes can track the aiming signal for not dramatic changes.
The proposed algorithm can track the slowly varying graph signal along with time.

\subsubsection{Parameters $\beta, \mu$, and $\Delta$}
\begin{figure}[!h]
\begin{center}
\includegraphics[width=\figurewidth]{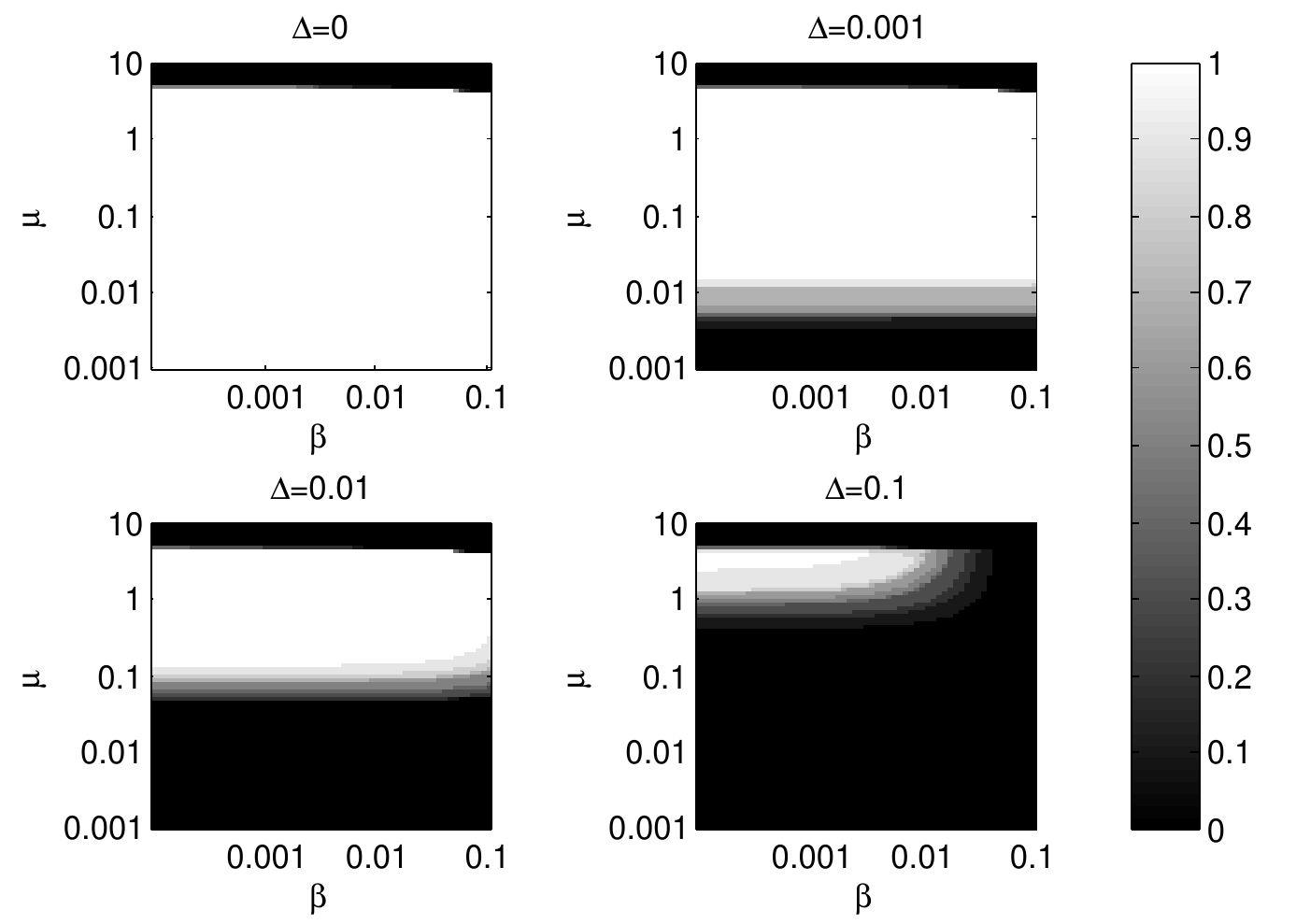}
\caption{The probability of convergence for different choices of $\mu$ and $\beta$ in time-invariant and time-varying cases.}
\label{exp12}
\end{center}
\end{figure}

In this experiment, the regions for the parameters $\beta$ and $\mu$ that guarantee the convergence of DLSR are plotted in Fig. \ref{exp12} for different $\Delta$, which describes the varying rate of time-varying signals.
The experiment results show that for time-varying case the algorithm is not convergent if the stepsize is too large or too small. If the $\mu$ is too small, the estimation cannot track the varying signal. This will not happen for the time-invariant case ($\Delta=0$).

\subsection{Reconstruction of Time-Invariant Signal}
\subsubsection{Convergence Performance}
\begin{figure}[!h]
\begin{center}
\includegraphics[width=\figurewidth]{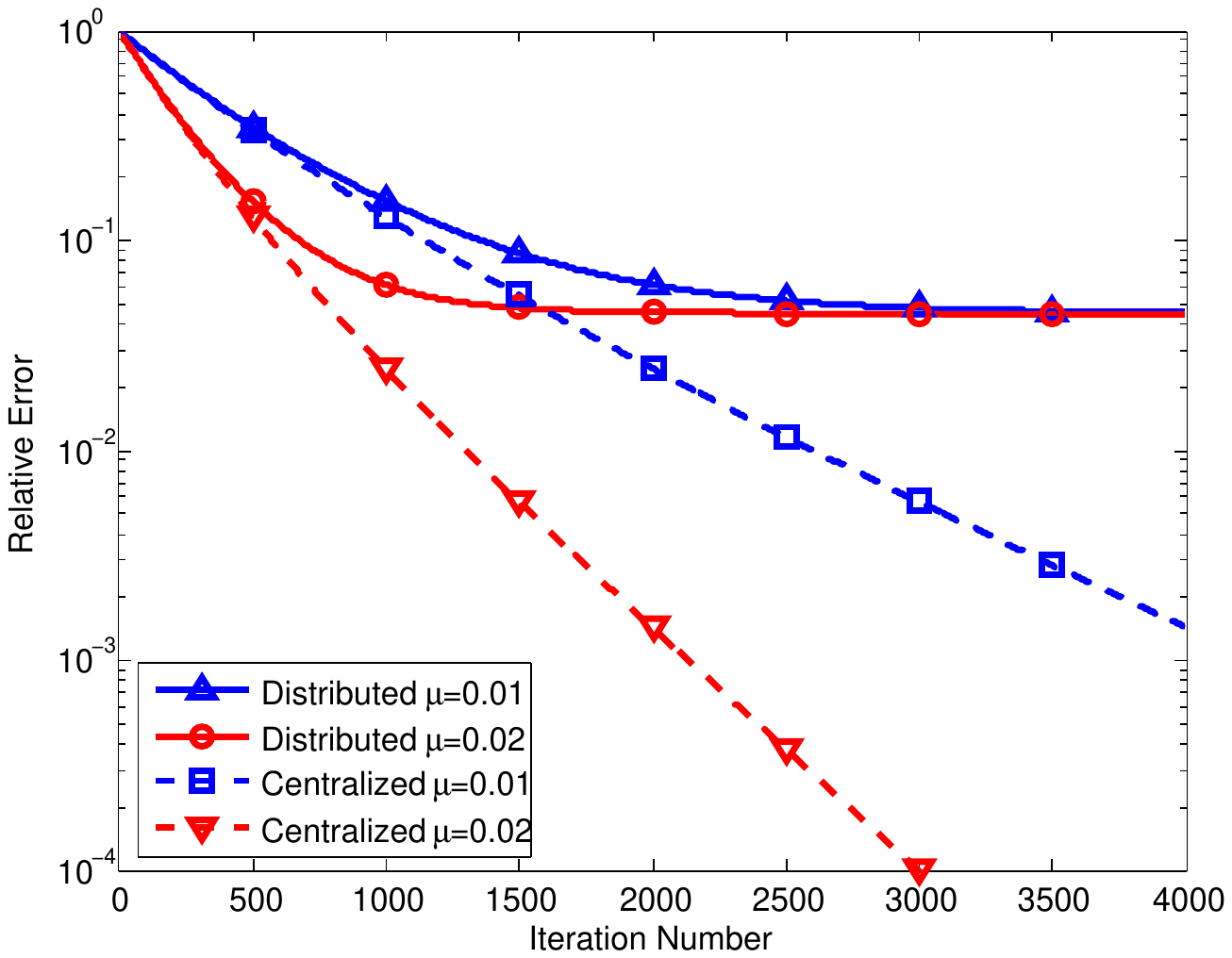}
\caption{The convergence curves of distributed and centralized algorithms with different constant stepsizes, where the decay factor is fixed $\beta=0.01$.}
\label{exp1}
\end{center}
\end{figure}

In this experiment, DLSR is used to reconstruct time-invariant signals. The convergence curves of distributed and centralized algorithms with constant stepsizes $\mu=0.01$ and $\mu=0.02$ are illustrated in Fig. \ref{exp1}, with the decay factor $\beta=0.01$. The centralized algorithm uses fresh data from the sampled nodes, while the distributed algorithm uses data with transmission delay. It is easy to see that a larger stepsize results in a faster convergence.
The bias caused by $\beta$ can be seen in the convergence curves of distributed algorithms, while the error of centralized algorithm shrinks exponentially to zero with no bias.

\subsubsection{In-Band and Out-of-Band Errors}

\begin{figure}[t]
\begin{center}
\includegraphics[width=\figurewidth]{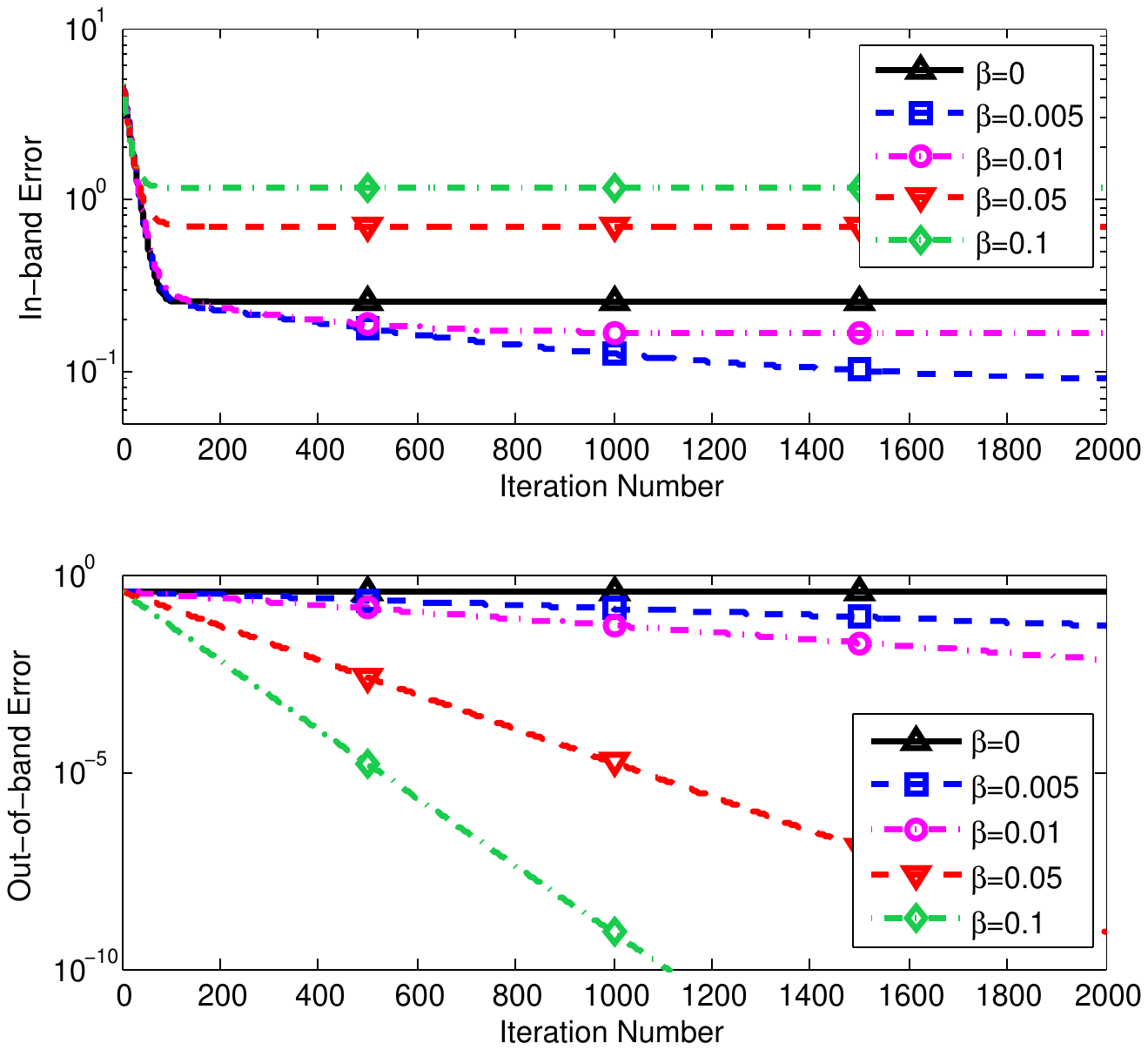}
\caption{The in-band and out-of-band errors for different $\beta$ if the initial value has out-of-band energy. If $\beta=0$ the out-of-band energy cannot be eliminated as the iteration goes, and it also leads to a relatively larger in-band error. For $\beta>0$, a larger $\beta$ will lead to a faster shrinkage of the out-of-band error, and a larger steady-state in-band error.}
\label{exp4}
\end{center}
\end{figure}

As proved in Corollary \ref{TimeInvariant}, both the in-band and out-of-band errors will decrease into a small bound as the iteration goes for $\beta>0$.
In this experiment, we set the initial value ${\bf f}^{(0)}$ with about $10\%$ of out-of-band energy and conduct the iteration with different decay factors $\beta=0, 0.005, 0.01, 0.05$, and $0.1$.
The stepsize is chosen as $\mu=0.2$. The in-band and out-of-band errors are illustrated in Fig. \ref{exp4}.
The experiment result shows the necessity of the decay factor. Although there is no bias for $\beta=0$, the out-of-band energy cannot be eliminated as the iteration goes.
It should be noted that the curve for $\beta=0$ is not comparable with the others because the out-of-band energy also affects the in-band error according to Remark \ref{rem2}.
Since the out-of-band energy cannot be eliminated, it also leads to a relatively larger in-band error for $\beta=0$.
For $\beta>0$, it can be seen from Fig. \ref{exp4} that even though the initial value has out-of-band energy, it will shrink towards zero along with the iteration.
A larger $\beta$ will lead to a faster shrinkage of the out-of-band error, and a larger steady-state in-band error.

\subsubsection{Constant Parameters $\beta$ and $\mu$}
\begin{figure}[t]
\begin{center}
\includegraphics[width=\figurewidth]{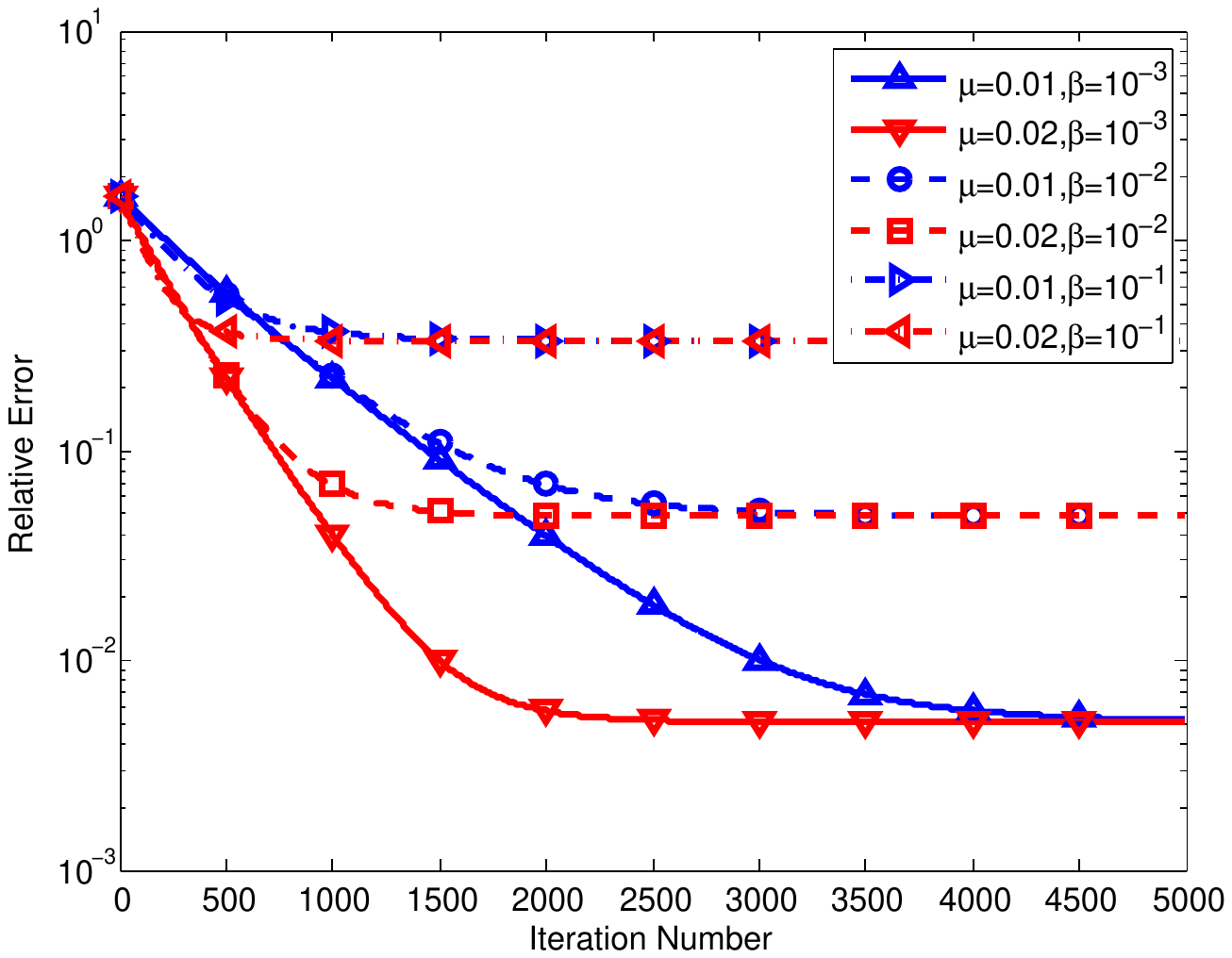}
\caption{Convergence curves for different choices of $\beta$ and $\mu$. The decay factor $\beta$ mainly determines the steady-state error and the stepsize $\mu$ mainly determines the convergence rate.}
\label{exp5}
\end{center}
\end{figure}

The convergence curves for different choices of constant $\beta$ and $\mu$ are illustrated in Fig. \ref{exp5}.
As analyzed above, the convergence rate is mainly determined by the stepsize $\mu$.
The steady-state error is composed of two parts, the bias and the steady-state error introduced by the constant stepsize.
The latter is relatively small compared with the former, which is mainly determined by the decay factor $\beta$.
It is obvious that a smaller $\beta$ will lead to a smaller steady-state error.

\begin{figure}[!h]
\begin{center}
\includegraphics[width=\figurewidth]{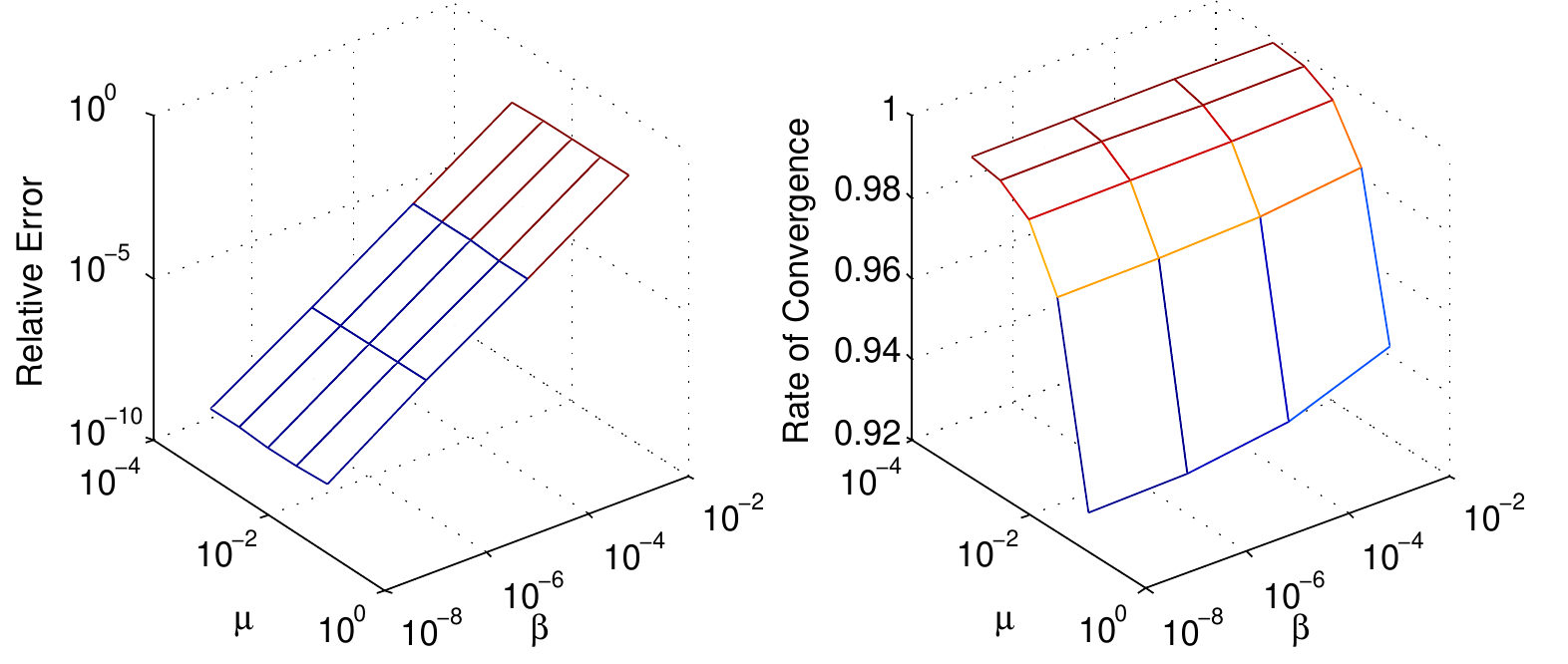}
\caption{The steady-state errors and convergence rates for different choices of $\beta$ and $\mu$.}
\label{exp9}
\end{center}
\end{figure}

The steady-state errors and convergence rates for different choices of $\beta$ and $\mu$ are plotted in Fig. \ref{exp9}.
For fixed $\beta$, the steady-state error varies little with $\mu$.
It shows that the bias, which is determined by $\beta$, is dominant in the total error, while $\mu$ influences the total error little.
Since there is bias in the convergence, the convergence rate is approximately calculated as $(\|{\bf f}^{(m)}-{\bf f}_*\|/\|{\bf f}^{(0)}-{\bf f}_*\|)^{1/m}$, where $m$ is the number of iterations
when the error reaches $1.2$ times the steady-state error.
The rate of convergence is smaller for larger $\mu$, which means it converges faster.

\subsubsection{Diminishing Parameters $\mu_k$ and $\beta_k$}

\begin{figure}[!h]
\begin{center}
\includegraphics[width=\figurewidth]{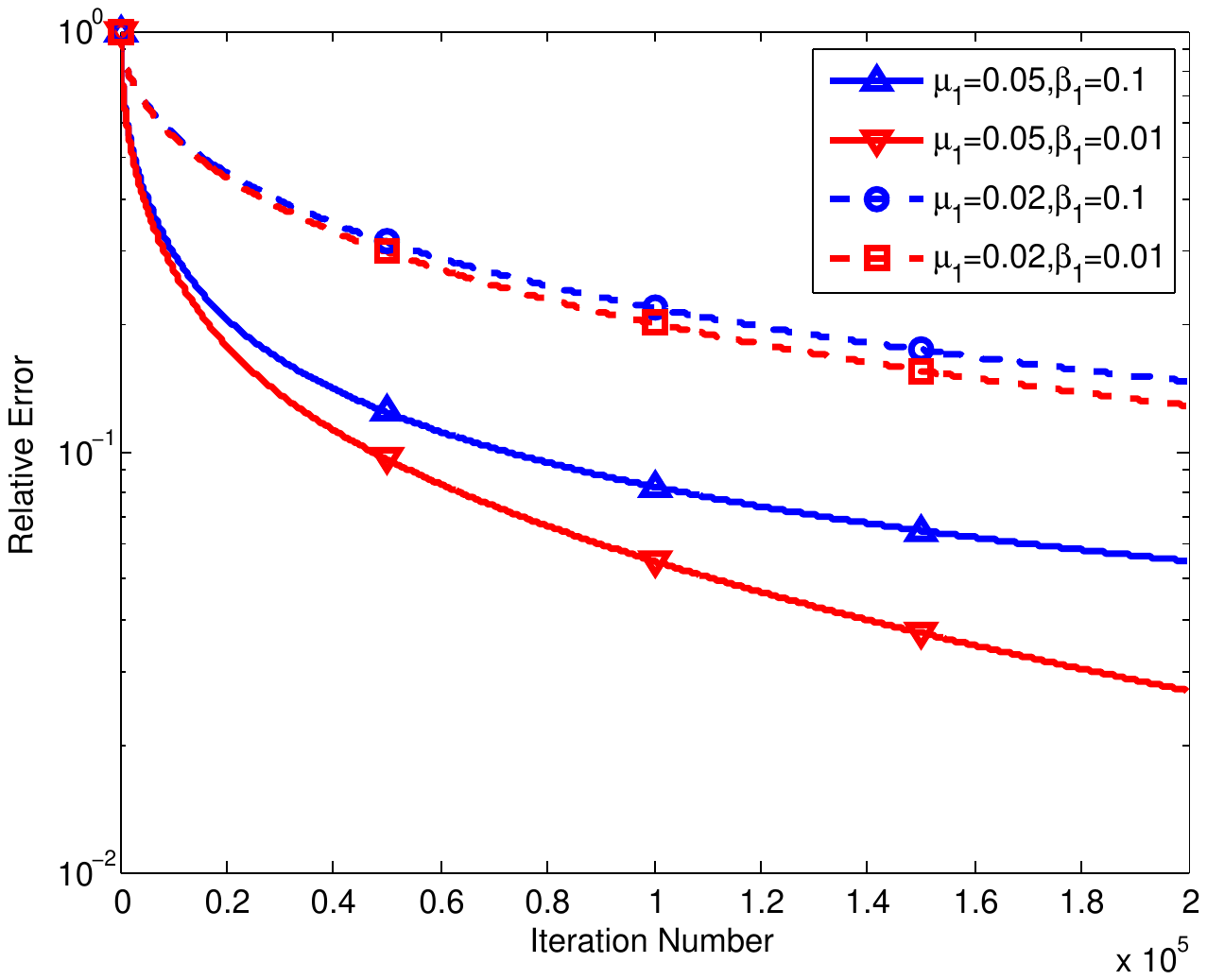}
\caption{The convergence curves for diminishing stepsizes and decay factors in Proposition \ref{pro3}.}
\label{exp10}
\end{center}
\end{figure}

An experiment for diminishing stepsizes and decay factors is conducted and the convergence curves are shown in Fig. \ref{exp10}.
The stepsizes are chosen as $\mu_k=\mu_1/\sqrt{k}$ with $\mu_1=0.05$ or $0.02$.
The decay factor are $\beta_k=\beta_1/\sqrt[4]{k}$ with $\beta_1=0.1$ or $0.01$.
All the curves decline along the iteration.
Among the four curves, it can be seen that a larger $\mu_1$ and a smaller $\beta_1$ may lead to a faster convergence.

\subsection{Experiments with Real Data}

\begin{figure}[t]
\begin{center}
\includegraphics[width=\figurewidth]{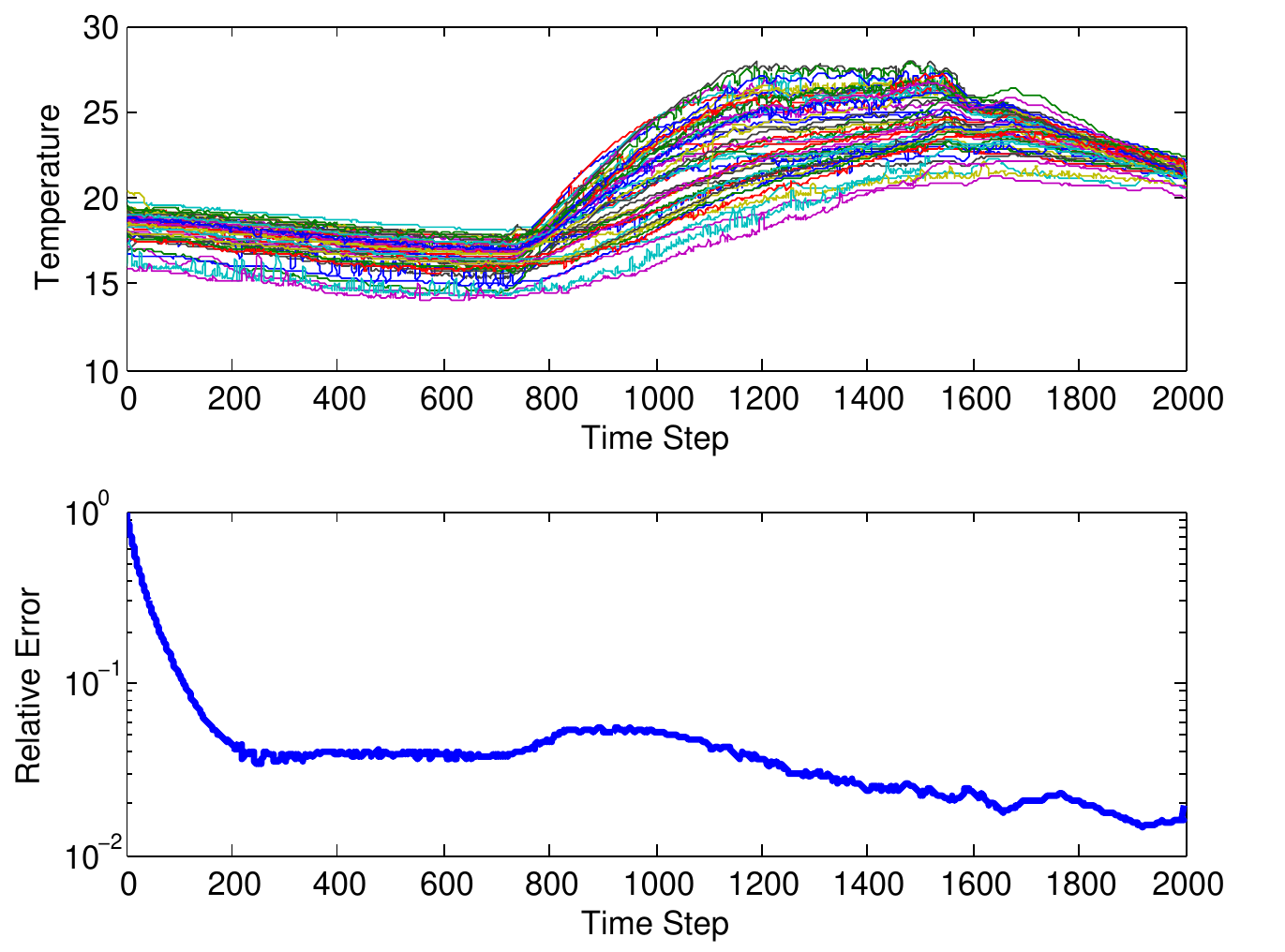}
\caption{The temperature of each node in the sensor network data and the relative error of DLSR.}
\label{RealData}
\end{center}
\end{figure}

The sensor network data of Intel Berkeley Research Lab \cite{IntelData} is used in this experiment.
The data is collected from $54$ sensors in the lab and sampled every $30$ seconds from February 28th, 2004, including temperature, humidity, light, and voltage.
In our experiment, the graph signal is composed of the temperature of the sensors.
We extract the data from 01:06:15 to 17:56:15 on February 28th, 2004, during which time there is less missing data.
Taking time and space smoothness into consideration, the missing data is interpolated by conducting the MATLAB function scatteredInterpolant with all the existing data.
Then the completed data is regarded as the original time-varying graph signal.
The graph is established by the $4$-nearest neighbors of the positions of the sensors, and the weights are inversely proportional to the square of geometric distance.
We randomly choose $20$ sensors and reconstruct the temperature of the other sensors.
By selecting $\mu=0.1$ and $\beta=10^{-3}$, the time-varying graph signal is reconstructed by DLSR.
The temperature of each node and the relative error are illustrated in Fig. \ref{RealData}.
The steady-state relative error is around $3\%$, which verifies the effectiveness of DLSR.

\section{Conclusion}

In this paper, the distributed least square reconstruction algorithm (DLSR) is proposed to estimate and track the unobserved data of a time-varying graph signal adaptively.
The low-frequency and high-frequency components of the recovered signals are theoretically proved to converge, respectively, to their true values.
The out-of-band energy caused by node-to-node transmission delay can be eliminated by using the decay factor, which introduces a controllable bias.
The expression of the overall error bound is given as a function of the stepsize and decay factor,  and can be made arbitrarily small.
Numerical experiments on both synthetic and real world data verify the performance of the proposed algorithm and show that DLSR is able to track slowly varying graph signals adaptively.

\section{Appendix}

\subsection{The Proof of Proposition \ref{TimeVariant}}\label{Proof}

First, besides the in-band error and out-of-band error defined in \eqref{eq:eofk} and \eqref{eq:epofk}, two sequences of quantities are introduced as
\begin{align}
\delta^{(k)}&=\left\|{\bf f}^{(k)}-{\bf f}^{(k-1)}\right\|,\label{eq:deltaofk}\\
\eta^{(k)}&=\left\|\sum_{u\in\mathcal{S}}\left(f^{(k)}(u){\bf I}_N-{\bf F}_u^{(k)}\right)\mathcal{P}_{\omega}\bm{\delta}_u\right\|.\label{eq:etaofk}
\end{align}
Then we will prove the following inequalities, where the proofs are postponed to the end of this subsection.
\begin{align}
e^{(k+1)}\le&(1-\mu\beta-\mu A)e^{(k)}+\mu \|{\bf T}\|e_+^{(k)}+\mu\eta^{(k)}+\left(\sqrt{N}+\mu |\mathcal{S}|\tau\right)\Delta,\label{IneqE}\\
e_+^{(k+1)}\le&(1-\mu\beta)e_+^{(k)}+\mu\eta^{(k)}+\left(\sqrt{N}+\mu |\mathcal{S}|\tau\right)\Delta,\label{IneqEplus}\\
\eta^{(k)}\le& \sqrt{|\mathcal{S}|}\sum_{i=0}^{\tau-1}\delta^{(k-i)},\label{IneqEta}\\
\delta^{(k)}\le& \mu \left((\beta+\|{\bf T}\|)\left(e^{(k-1)}+e_+^{(k-1)}\right)+\eta^{(k-1)}\right)
+\mu |\mathcal{S}|\tau\Delta,\label{IneqDelta}
\end{align}

Plugging \eqref{IneqDelta} into \eqref{IneqEta}, we have
\begin{align}\label{Etak}
\eta^{(k)}\le \mu\left(C+|\mathcal{S}|^{\frac{3}{2}}\tau^2\Delta\right),
\end{align}
where $C$ is defined as (\ref{DefC}).

By plugging \eqref{Etak} into \eqref{IneqEplus} and \eqref{IneqE}, respectively, we have
\eqref{TVOutBandErr} and \eqref{TVInBandErr} readily. As a consequent, if we could demonstrate that $\{e^{(i)}\}$, $\{e_+^{(i)}\}$, and $\left\{\eta^{(i)}\right\}$ are bounded by some constants, respectively, the proof of Proposition \ref{TimeVariant} will be closed. In what follows, we will prove this by mathematical induction.

For $i=1$, these quantities are obviously bounded. We will then prove that if the preceding $k-1$ items of $\left\{e^{(i)}\right\}$, $\{e_+^{(i)}\}$, and $\left\{\eta^{(i)}\right\}$ have respective bounds of $B_e$, $B_{e_+}$, and $B_{\eta}$, their $k$th items are also bounded by the same limits.

\begin{enumerate}
\item
According to \eqref{Etak}, $\eta^{(k)}$ is bounded by $B_{\eta}$.
\item
Plugging $e_+^{(k-1)}\le B_{e_+}$ in \eqref{TVOutBandErr}, we have
$$
e_+^{(k)}\le(1-\mu\beta)B_{e_+}+\mu^2C+M(\mu)\Delta,
$$
where $M(\mu)$ is defined in \eqref{eq:Mofmu}. We then have $e_+^{(k)}\le B_{e_+}$ if
\begin{equation}\label{IneqOfMu}
\left(C+|\mathcal{S}|^{\frac{3}{2}}\tau^2\Delta\right)\mu^2+\left(|\mathcal{S}|\tau\Delta-\beta B_{e_+}\right)\mu+\sqrt{N}\Delta\le0.
\end{equation}
One may notice that the inequality (\ref{IneqOfMu}) has solutions for $\mu$, if and only if the following inequality holds for $\Delta$,
$$
\left(|\mathcal{S}|\tau\Delta-\beta B_{e_+}\right)^2-4\sqrt{N}\Delta\left(C+|\mathcal{S}|^{\frac{3}{2}}\tau^2\Delta\right)\ge 0,
$$
which is equivalent to
\begin{align}
&|\mathcal{S}|^{\frac{3}{2}}\tau^2\left(4N^{\frac{1}{2}}-|\mathcal{S}|^{\frac{1}{2}}\right)\Delta^2+
\left(2|\mathcal{S}|\tau\beta B_{e_+} +4N^{\frac{1}{2}} C\right)\Delta\le \beta^2 B_{e_+}^2.\label{eq:kappaconstraint}
\end{align}
Because its left hand side is an increasing function of $\Delta$, the inequality \eqref{eq:kappaconstraint} is satisfied if
$\Delta\le\Delta_{\text{max}}$, where $\Delta_{\text{max}}$ can be solved from \eqref{eq:kappaconstraint}.

Then from (\ref{IneqOfMu}) the range of $\mu$ can be determined as $\mu_{\text{min}}\le\mu\le\mu_{\text{max}}$, where both
$\mu_{\text{min}}$ and $\mu_{\text{max}}$ are related to $\Delta$.
\item
Plugging $e^{(k-1)}\le B_{e}$ and $e_+^{(k-1)}\le B_{e_+}$ into (\ref{IneqE}), we have
\begin{align}
e^{(k)}\le (1-\mu\beta-\mu A)B_e+\mu\|{\bf T}\|B_{e_+}
+\mu^2C+M(\mu)\Delta\nonumber.
\end{align}
Using \eqref{eq:additionalcondition}, we can also obtain that $e^{(k)}\le B_{e}$ if (\ref{IneqOfMu}) is satisfied.
\end{enumerate}
Consequently, $\{e^{(k)}\}$, $\{e_+^{(k)}\}$,  and $\{\eta^{(k)}\}$ are bounded, and then Proposition \ref{TimeVariant} is proved.

To end this subsection, we will prove \eqref{IneqE}, \eqref{IneqEplus}, \eqref{IneqEta}, and \eqref{IneqDelta}. To simplify the expression, we introduce two vectors to denote the misalignment of estimated signal and the increment of true signal by
\begin{align}
{\bf d}^{(k)} &= {\bf f}^{(k)}-\tilde{\bf f}_*^{(k)},\label{eq:ferror}\\
{\bf c}^{(k)} &= \tilde{\bf f}_*^{(k)}-\tilde{\bf f}_*^{(k-1)},\label{eq:finc}
\end{align}
and two diagonal matrices to denote the errors at node $u$ caused by delayed true signal and estimated signal by
\begin{align}
{\bf E}_{*u}^{(k)} &= f_*^{(k)}(u){\bf I}_N-{\bf F}_{*u}^{(k)},\label{eq:Etrueerror}\\
{\bf E}_{u}^{(k)} &= f^{(k)}(u){\bf I}_N-{\bf F}_{u}^{(k)},\label{eq:Eerror}
\end{align}

\subsubsection{The Proof of (\ref{IneqE}) and (\ref{IneqEplus})}

According to (\ref{DistIter}), (\ref{DefS}), and (\ref{FStar}), we have
\begin{align}\label{FkMinusFstar}
{\bf d}^{(k+1)}
=&(1-\mu\beta){\bf d}^{(k)}-{\bf c}^{(k+1)}-\mu\beta\tilde{\bf f}_*^{(k)}
+\mu\sum_{u\in\mathcal{S}}({\bf F}_{*u}^{(k)}-{\bf F}_u^{(k)})\mathcal{P}_{\omega}\bm{\delta}_u\nonumber\\
=&(1-\mu\beta){\bf d}^{(k)}-{\bf c}^{(k+1)}-\mu{\bf T}{\bf d}^{(k)}
+\mu{\bf T}({\bf f}^{(k)}-{\bf f}_*^{(k)})+\mu\sum_{u\in\mathcal{S}}({\bf F}_{*u}^{(k)}-{\bf F}_u^{(k)})\mathcal{P}_{\omega}\bm{\delta}_u\nonumber\\
=&{\bf Q}\left(\mathcal{P}_{\omega}{\bf d}^{(k)}+\mathcal{P}_{\omega_+}{\bf d}^{(k)}\right)-{\bf c}^{(k+1)}
+\mu\sum_{u\in\mathcal{S}}{\bf E}_u^{(k)}\mathcal{P}_{\omega}\bm{\delta}_u-\mu\sum_{u\in\mathcal{S}}{\bf E}_{*u}^{(k)}\mathcal{P}_{\omega}\bm{\delta}_u,
\end{align}
where
$$
	{\bf Q} = (1-\mu\beta){\bf I}_N-\mu{\bf T}.
$$

Considering the definition of ${\bf T}$ in (\ref{DefS}), for any ${\bf f}$, we have
\begin{align}
\mathcal{P}_{\omega}{\bf Q}\mathcal{P}_{\omega}{\bf f}&={\bf Q}\mathcal{P}_{\omega}{\bf f},\label{LL}\\
\mathcal{P}_{\omega}{\bf Q}\mathcal{P}_{\omega_+}{\bf f}&=-\mu{\bf T}\mathcal{P}_{\omega_+}{\bf f},\label{LH}\\
\mathcal{P}_{\omega_+}{\bf Q}\mathcal{P}_{\omega}{\bf f}&={\bf 0},\label{HL}\\
\mathcal{P}_{\omega_+}{\bf Q}\mathcal{P}_{\omega_+}{\bf f}&=(1-\mu\beta)\mathcal{P}_{\omega_+}{\bf f}.\label{HH}
\end{align}

Therefore, according to (\ref{LL}) and (\ref{LH}), the low-frequency part of (\ref{FkMinusFstar}) is
\begin{align}\label{Pwf-f*}
\mathcal{P}_{\omega}{\bf d}^{(k+1)}
={\bf Q}\mathcal{P}_{\omega}{\bf d}^{(k)}-\mu{\bf T}\mathcal{P}_{\omega_+}{\bf d}^{(k)}-\mathcal{P}_{\omega}{\bf c}^{(k+1)}
+\mu\mathcal{P}_{\omega}\sum_{u\in\mathcal{S}}{\bf E}_u^{(k)}\mathcal{P}_{\omega}\bm{\delta}_u
-\mu\mathcal{P}_{\omega}\sum_{u\in\mathcal{S}}{\bf E}_{*u}^{(k)}\mathcal{P}_{\omega}\bm{\delta}_u.
\end{align}

Since the frame bound of $\{\mathcal{P}_{\omega}\bm{\delta}_u\}_{u\in\mathcal{S}}$ satisfies $A\le B\le1$,
if we choose a stepsize $\mu$ satisfying $\mu<1/(\beta+A)$,
the assumption $A{\bf I}_N\preceq {\bf T}\preceq B{\bf I}_N$ implies, note that $\mathcal{P}_{\omega}{\bf d}^{(k)}\in PW_{\omega}(\mathcal{G})$,
$$
\|{\bf Q}\|\le1-\mu\beta-\mu A.
$$
According to (\ref{kappa}),
$$
\left\|\mathcal{P}_{\omega}{\bf c}^{(k+1)}\right\|\le\left\|{\bf c}^{(k+1)}\right\|\le\sqrt{N}\Delta.
$$
The definition of ${\bf F}_{*u}^{(k)}$ implies
$$
-\tau\Delta{\bf I}_N\preceq {\bf E}_{*u}^{(k)}\preceq \tau\Delta{\bf I}_N,
$$
and then the last term of (\ref{Pwf-f*}) is bounded by
$$
\left\|\mathcal{P}_{\omega}\left(\sum_{u\in\mathcal{S}}{\bf E}_{*u}^{(k)}\mathcal{P}_{\omega}\bm{\delta}_u\right)\right\|\le|\mathcal{S}|\tau\Delta.
$$

Taking the norm of (\ref{Pwf-f*}) and combining the above inequalities, the inequality (\ref{IneqE}) is obtained.

According to (\ref{HL}) and (\ref{HH}), the high-frequency part of (\ref{FkMinusFstar}) is
\begin{align}
\mathcal{P}_{\omega_+}{\bf d}^{(k+1)}
=&(1-\mu\beta)\mathcal{P}_{\omega_+}{\bf d}^{(k)}-\mathcal{P}_{\omega_+}{\bf c}^{(k+1)}
+\mu\mathcal{P}_{\omega_+}\sum_{u\in\mathcal{S}}{\bf E}_u^{(k)}\mathcal{P}_{\omega}\bm{\delta}_u
-\mu\mathcal{P}_{\omega_+}\sum_{u\in\mathcal{S}}{\bf E}_{*u}^{(k)}\mathcal{P}_{\omega}\bm{\delta}_u.\nonumber
\end{align}
Following the samilar approach of proving (\ref{IneqE}), the inequality (\ref{IneqEplus}) is proved.
\subsubsection{The Proof of (\ref{IneqEta})}
By the definition of ${\bf F}_u^{(k)}$,
\begin{align}
\left\|{\bf E}_u^{(k)}\mathcal{P}_{\omega}\bm{\delta}_u\right\|^2
=&\sum_v\left(\left|f^{(k)}(u)-f^{(k-\tau(u,v))}(u)\right|^2\left|(\mathcal{P}_{\omega}\bm{\delta}_u)(v)\right|^2\right)\nonumber\\
\le&\sum_v\left|(\mathcal{P}_{\omega}\bm{\delta}_u)(v)\right|^2\left(\sum_{i=0}^{\tau-1}
\left|f^{(k-i)}(u)-f^{(k-i-1)}(u)\right|\right)^2\nonumber\\
\le&\left(\sum_{i=0}^{\tau-1}\left|f^{(k-i)}(u)-f^{(k-i-1)}(u)\right|\right)^2\nonumber,
\end{align}
where
$$
\sum_v\left|(\mathcal{P}_{\omega}\bm{\delta}_u)(v)\right|^2=\left\|\mathcal{P}_{\omega}\bm{\delta}_u\right\|^2\le 1.
$$

Therefore,
\begin{align}
\eta^{(k)}\le&\sum_{u\in\mathcal{S}}\left\|{\bf E}_u^{(k)}\mathcal{P}_{\omega}\bm{\delta}_u\right\|\nonumber\\
\le&\sum_{u\in\mathcal{S}}\sum_{i=0}^{\tau-1}
\left|f^{(k-i)}(u)-f^{(k-i-1)}(u)\right|\nonumber\\
\le&\sum_{i=0}^{\tau-1}\sqrt{|\mathcal{S}|}\left(\sum_{u\in\mathcal{S}}\left|f^{(k-i)}(u)-f^{(k-i-1)}(u)\right|^2\right)^{\frac{1}{2}}\nonumber\\
\le&\sqrt{|\mathcal{S}|}\sum_{i=0}^{\tau-1}\left\|{\bf f}^{(k-i)}-{\bf f}^{(k-i-1)}\right\|,\nonumber
\end{align}
which is (\ref{IneqEta}).

\subsubsection{The Proof of (\ref{IneqDelta})}
According to (\ref{DistIter}), (\ref{DefS}), and (\ref{FStar}),
\begin{align}\label{PFkMinusPFk}
{\bf f}^{(k)}-{\bf f}^{(k-1)}
=&-\mu\beta{\bf f}^{(k-1)}+\mu\sum_{u\in\mathcal{S}}\left({\bf F}_{*u}^{(k-1)}-{\bf F}_u^{(k-1)}\right)\mathcal{P}_{\omega}\bm{\delta}_u\nonumber\\
=&-\mu\beta{\bf d}^{(k-1)}-\mu{\bf T}{\bf d}^{(k-1)}
+\mu\sum_{u\in\mathcal{S}}{\bf F}_{*u}^{(k-1)}\mathcal{P}_{\omega}\bm{\delta}_u
-\mu\sum_{u\in\mathcal{S}}{\bf F}_u^{(k-1)}\mathcal{P}_{\omega}\bm{\delta}_u\nonumber\\
=&-\mu(\beta{\bf I}_N+{\bf T}){\bf d}^{(k-1)}
+\mu\sum_{u\in\mathcal{S}}{\bf E}_u^{(k-1)}\mathcal{P}_{\omega}\bm{\delta}_u
-\mu\sum_{u\in\mathcal{S}}{\bf E}_{*u}^{(k-1)}\mathcal{P}_{\omega}\bm{\delta}_u.
\end{align}

Taking the norm of (\ref{PFkMinusPFk}), the inequality (\ref{IneqDelta}) is obtained.

\subsection{The Proof of Proposition \ref{pro3}}\label{proof2}
According to the proof of Proposition \ref{TimeVariant}, setting $\Delta=0$ and using $\|{\bf T}\|\le 1$ and $\beta_k\le \beta_1$, the inequalities (\ref{IneqE})-(\ref{IneqDelta}) for variant $\{\mu_k\}$ and $\{\beta_k\}$ becomes
\begin{align}
e^{(k)}\le&(1-\mu_{k}\beta_{k}-\mu_{k} A)e^{(k-1)}+\mu_{k}e_+^{(k-1)}+\mu_{k}\eta^{(k-1)}\label{IneqEk}\\
e_+^{(k)}\le&(1-\mu_{k}\beta_{k})e_+^{(k-1)}+\mu_{k}\eta^{(k-1)}\label{IneqEplusk}\\
\eta^{(k)}\le& \sqrt{|\mathcal{S}|}\sum_{i=0}^{\tau-1}\delta^{(k-i)}\label{IneqEtak}\\
\delta^{(k)}\le& \mu_k \left((\beta_1+1)(e^{(k-1)}+e_+^{(k-1)})+\eta^{(k-1)}\right)\label{IneqDeltak}.
\end{align}

Similar to the proof of Proposition \ref{TimeVariant}, it is easy to see that $\{e^{(i)}\}$, $\{e_+^{(i)}\}$ and $\{\eta^{(i)}\}$ are bounded by constants $B_e$, $B_{e_+}$ and $B_{\eta}$ for $\mu_k=\mu_1/\sqrt{k}$.
Plugging (\ref{IneqEtak}) and (\ref{IneqDeltak}) into (\ref{IneqEplusk}), we have
\begin{equation}\label{mukmutau}
e_+^{(k)}\le(1-\mu_{k}\beta_{k})e_+^{(k-1)}+C'\mu_k\mu_{k-\tau},
\end{equation}
where $C'=\tau\sqrt{|\mathcal{S}|}\left(2(B_e+B_{e_+})+B_{\eta}\right)$.
For $\mu_k=\mu_1/\sqrt{k}$ and $\beta_k=\beta_1/\sqrt[4]{k}$, if $e_+^{(k-1)}\le L_+/\sqrt[4]{k-1}$ is satisfied for a constant $L_+$,
according to (\ref{mukmutau}), $e_+^{(k)}\le L_+/\sqrt[4]{k}$ as long as the following inequality is satisfied,
$$
\left(1-\frac{\mu_1\beta_1}{\sqrt{k}\sqrt[4]{k}}\right)\frac{L_+}{\sqrt[4]{k-1}}+C'\frac{\mu_1^2}{\sqrt{k}\sqrt{k-\tau}}\le\frac{L_+}{\sqrt[4]{k}}.
$$
The inequality above is equivalent to
\begin{equation}\label{Lplus}
\frac{\mu_1^2C'}{L_+}\frac{\sqrt[4]{k}\sqrt[4]{k-1}}{\sqrt{k-\tau}}+\frac{\sqrt{k}}{(\sqrt[4]{k}+\sqrt[4]{k-1})(\sqrt{k}+\sqrt{k-1})}\le \mu_1\beta_1.
\end{equation}
Because the first term of the left side approaches $\mu_1^2C'/L_+$ and the second term approaches $0$ when $k$ is large enough, by selecting a constant $L_+$ appropriately, (\ref{Lplus}) is established and then we have $e_+^{(k)}\le L_+/\sqrt[4]{k}$.

Plugging (\ref{IneqEtak}) and (\ref{IneqDeltak}) into (\ref{IneqEk}), we have
$$
e^{(k)}\le(1-\mu_{k}\beta_{k}-\mu_{k} A)e^{(k-1)}+\mu_{k}e_+^{(k-1)}+C'\mu_k\mu_{k-\tau}.
$$
If $e^{(k-1)}\le L/\sqrt[4]{k-1}$ is satisfied for a constant $L$,
$e^{(k)}\le L/\sqrt[4]{k}$ as long as the following inequality is satisfied,
\begin{align}
\left(1-\frac{\mu_1}{\sqrt{k}}\left(\frac{\beta_1}{\sqrt[4]{k}}+A\right)\right)\frac{L}{\sqrt[4]{k-1}}
+\frac{\mu_1}{\sqrt{k}}\frac{L_+}{\sqrt[4]{k-1}}
+C'\frac{\mu_1^2}{\sqrt{k}\sqrt{k-\tau}}\le\frac{L}{\sqrt[4]{k}},\nonumber
\end{align}
which is equivalent to
\begin{align}\label{L}
&\frac{\mu_1L_+}{L}+\frac{\mu_1^2C'}{L}\frac{\sqrt[4]{k-1}}{\sqrt{k-\tau}}+\frac{\sqrt[4]{k}}{(\sqrt[4]{k}+\sqrt[4]{k-1})(\sqrt{k}+\sqrt{k-1})}
\le \mu_1\left(A+\frac{\beta_1}{\sqrt[4]{k}}\right).
\end{align}
Both the second and third terms of (\ref{L}) approach $0$ when $k$ is large enough.
By selecting a constant $L$ appropriately, (\ref{L}) is established and then we have $e^{(k)}\le L/\sqrt[4]{k}$.

Based on the analysis above, we have
\begin{align}
\|{\bf f}^{(k)}-{\bf f}_*\|&\le\frac{\beta_k}{\beta_k+A}\|{\bf f}_*\|+e_+^{(k)}+e^{(k)}\nonumber\\
&\le\left(\frac{\beta_1}{A}\|{\bf f}_*\|+L_++L\right)\frac{1}{\sqrt[4]{k}},
\end{align}
when $k$ is large, and Proposition \ref{pro3} is proved.


\footnotesize
\begin{thebibliography}{1}
\bibitem{shuman_emerging_2013}
D. I. Shuman, S. K. Narang, P. Frossard, A. Ortega, and P. Vandergheynst, ``The
  emerging field of signal processing on graphs: Extending high-dimensional
  data analysis to networks and other irregular domains,'' \emph{IEEE Signal
  Process. Mag.}, vol.~30, no.~3, pp. 83-98, 2013.

\bibitem{sandryhaila_discrete_2013}
A. Sandryhaila, and J. M. F. Moura, ``Discrete signal processing on graphs,'' \emph{IEEE Trans. Signal Process.}, vol. 61, no. 7, pp. 1644-1656, 2013.

\bibitem{zhu_graph_2012}
X. Zhu and M. Rabbat, ``Graph spectral compressed sensing for sensor networks,'' in \emph{Proc. 37th IEEE Int. Conf. Acoust., Speech, Signal Process. (ICASSP)}, 2012, pp. 2865-2868.

\bibitem{narang_graph_2012}
S. K. Narang, Y. H. Chao, and A. Ortega, ``Graph-wavelet filterbanks for edge-aware image processing,'' in \emph{ Proc. IEEE Stat.
Signal Process. Workshop (SSP'12)}, 2012, pp. 141-144.

\bibitem{narang_signal_2013}
S. K. Narang, A. Gadde, and A. Ortega, ``Signal processing techniques for interpolation in graph structured data,''
in \emph{Proc. 38th IEEE Int. Conf. Acoust., Speech, Signal Process. (ICASSP)}, 2013, pp. 5445-5449.

\bibitem{narang_localized_2013}
S. K. Narang, A. Gadde, E. Sanou, and A. Ortega, ``Localized iterative methods for interpolation in graph structured data,''
in \emph{Proc. 1st IEEE Global Conf. Signal and Inform. Process. (GlobalSIP)}, 2013, pp. 491-494.

\bibitem{anis_towards_2014}
A. Anis, A. Gadde, and A. Ortega, ``Towards a sampling theorem for signals on arbitrary graphs,'' in
\emph{Proc. 39th IEEE Int. Conf. Acoust., Speech, Signal Process. (ICASSP)}, 2014, pp. 3892-3896.

\bibitem{agaskar_aspectral_2013}
A. Agaskar, and Y. M. Lu, ``A spectral graph uncertainty principle,'' \emph{IEEE Trans. Inform. Theory}, vol. 59, no. 7, pp. 4338-4356, 2013.

\bibitem{chen_adaptive_2013}
S. Chen, A. Sandryhaila, J. M. F. Moura, and J. Kovacevic, ``Adaptive graph filtering: Multiresolution classification on graphs,'' in \emph{Proc. 1st IEEE Global Conf. Signal and Inform. Process. (GlobalSIP)}, pp. 427-430, 2013.

\bibitem{hammond_wavelets_2011}
D. K. Hammond, P. Vandergheynst, and R. Gribonval, ``Wavelets on graphs via spectral graph theory,''
\emph{Appl. Comput. Harmonic Anal.}, vol. 30, no. 2, pp. 129-150, 2011.

\bibitem{narang_perfect_2012}
S. K. Narang and A. Ortega, ``Perfect reconstruction two-channel wavelet filter-banks for graph structured data,'' \emph{IEEE Trans. Signal Process.}, vol. 60, no. 6, pp. 2786-2799, 2012.

\bibitem{zhu_approximating_2012}
X. Zhu and M. Rabbat, ``Approximating signals supported on graphs,'' in \emph{Proc. 37th IEEE Int. Conf. Acoust., Speech, Signal Process. (ICASSP)}, 2012, pp. 3921-3924.

\bibitem{shuman_aframework_2013}
D. I. Shuman, M. J. Faraji, and P. Vandergheynst, ``A framework for multiscale transforms on graphs,'' \emph{arXiv preprint arXiv:1308.4942}, 2013.

\bibitem{ekambaram_multiresolution_2013}
V. N. Ekambaram, G. C. Fanti, B. Ayazifar, and K. Ramchandran, ``Multiresolution graph signal processing via circulant structures,'' in \emph{Proc. IEEE Digital Signal Process., Signal Process. Educ. Meeting (DSP/SPE)}, 2013, pp. 112-117.

\bibitem{thanou_parametric_2013}
D. Thanou, D. I. Shuman, and P. Frossard, ``Parametric dictionary learning for graph signals,'' in \emph{Proc. 1st IEEE Global Conf. Signal and Inform. Process. (GlobalSIP)}, 2013, pp. 487-490.

\bibitem{liu_coarsening_2014}
P. Liu, X. Wang and Y. Gu, ``Coarsening graph signal with spectral invariance,'' in \emph{Proc. 39th IEEE Int. Conf. Acoust., Speech, Signal Process. (ICASSP)}, 2014, pp. 1075-1079.

\bibitem{liu_graphcoarsening_2014}
P. Liu, X. Wang, and Y. Gu, ``Graph signal coarsening: Dimensionality reduction in irregular domain,'' in \emph{Proc. 2nd IEEE Global Conf. Signal and Inform. Process. (GlobalSIP)}, 2014, pp. 966-970.

\bibitem{pesenson_sampling_2008}
I. Pesenson, ``Sampling in Paley-Wiener spaces on combinatorial graphs,''
\emph{Trans. Amer. Math. Soc.}, vol. 360, no. 10, pp. 5603-5627, 2008.

\bibitem{pesenson_variational_2009}
I. Pesenson, ``Variational splines and Paley-Wiener spaces on combinatorial graphs,''
\emph{Constructive Approximation}, vol. 29, pp. 1-21, 2009.

\bibitem{pesenson_sampling_2010}
I. Z. Pesenson, and M. Z. Pesenson, ``Sampling, filtering and sparse approximations on combinatorial graphs,''
\emph{J. Fourier Anal. and Applicat.}, vol. 16, no. 6, pp. 921-942, 2010.

\bibitem{wang_iterative_2014}
X. Wang, P. Liu, and Y. Gu, ``Iterative reconstruction of graph signal in low-frequency subspace,'' in \emph{Proc. 2nd IEEE Global Conf. Signal and Inform. Process. (GlobalSIP)}, 2014, pp. 611-615.

\bibitem{wang_localset_2014}
X. Wang, P. Liu, and Y. Gu, ``Local-set-based graph signal reconstruction,'' \emph{arXiv preprint arXiv:1410.3944}, 2014.

\bibitem{bar_multi_1995}
Y. Bar-Shalom and X. R. Li, \emph{Multitarget-Multisensor Tracking: Principles and Techniques}, Storrs, CT: University of Connecticut, 1995.

\bibitem{guestrin_distributed_2004}
C. Guestrin, P. Bodik, R. Thibaux, M. Paskin, and S Madden, ``Distributed regression: An efficient framework for modeling sensor network data,'' in \emph{Proc. 3rd Int. Symp. Inform. Process. in Sensor Networks (IPSN)}, 2004, pp. 1-10.

\bibitem{paskin_arobust_2005}
M. Paskin, C. Guestrin, and J. McFadden, ``A robust architecture for distributed inference in sensor networks,''
in \emph{Proc. 4th Int. Symp. Inform. Process. in Sensor Networks (IPSN)}, 2005, pp. 55-62.

\bibitem{xiao_ascheme_2005}
L. Xiao, S. Boyd, and S. Lall, ``A scheme for robust distributed sensor fusion based on average consensus,''
in \emph{Proc. 4th Int. Symp. Inform. Process. in Sensor Networks (IPSN)}, 2005, pp. 63-70.

\bibitem{schizas_concensus_2008}
I. D. Schizas, A. Ribeiro, and G. B. Giannakis, ``Consensus in ad hoc WSNs with noisy links-Part I: Distributed estimation of deterministic signals,'' \emph{IEEE Trans. Signal Process.}, vol. 56, no. 1, pp. 350-364, 2008.

\bibitem{olfati_distributed_2007}
R. Olfati-Saber, ``Distributed Kalman filtering for sensor networks,'' in \emph{Proc. 46th IEEE Conf. Decision and Control}, 2007, pp. 5492-5498.

\bibitem{cattivelli_rls_2008}
F. S. Cattivelli, C. G. Lopes, and A. H. Sayed, ``Diffusion recursive least-squares for distributed estimation over adaptive networks,''
\emph{IEEE Trans. Signal Process.}, vol. 56, no. 5, pp. 1865-1877, 2008.

\bibitem{cattivelli_lms_2010}
F. S. Cattivelli, and A. H. Sayed, ``Diffusion LMS strategies for distributed estimation,''
\emph{IEEE Trans. Signal Process.}, vol. 58, no. 3, pp. 1035-1048, 2010.

\bibitem{shuman_chebyshev_2011}
D. I. Shuman, P. Vandergheynst, and P. Frossard, ``Chebyshev polynomial approximation for distributed signal processing,'' in \emph{Proc. 7th Int. Conf. Distributed Computing in Sensor Syst. and Workshops (DCOSS)}, 2011, pp. 1-8.

\bibitem{chung_spectral_1997}
F. R. K. Chung, \emph{Spectral Graph Theory}, Amer. Math. Soc., 1997.

\bibitem{biyikoglu_laplacian_2007}
T. B\i y\i ko{\u g}lu, J. Leydold, and P. F. Stadler, ``Laplacian eigenvectors of
graphs,'' \emph{Lecture Notes in Mathematics}, vol. 1915, Springer, 2007.

\bibitem{Christensen_an_2002}
O. Christensen, \emph{An Introduction to Frames and Riesz Bases}, Springer, 2003.

\bibitem{POCS1}
W. Cheney, and A. Goldstein, ``Proximity maps for convex sets,'' \emph{Proc. Amer. Math. Soc.}, vol. 10, no. 3, pp. 448-450, 1959.

\bibitem{POCS2}
L. G. Gubin, B. T. Polyak, and E. V. Raik, ``The method of projections for finding the common point of convex sets,'' \emph{USSR Computational Mathematics and Mathematical Physics}, vol. 7, no. 6, pp. 1-24, 1967.

\bibitem{feichtinger_theory_1994}
H. G. Feichtinger, and K. Gr\"{o}chenig, ``Theory and practice of irregular sampling,''
\emph{Wavelets: Mathematics and Applications}, pp. 305-363, 1994.

\bibitem{grochenig_adiscrete_1993}
K. Gr\"{o}chenig, ``A discrete theory of irregular sampling,''
\emph{Linear Algebra and Its Applications}, vol. 193, pp. 129-150, 1993.

\bibitem{benedetto_irregular_1992}
J. J. Benedetto,``Irregular sampling and frames,''
\emph{Wavelets: A Tutorial in Theory and Applications}, vol. 2, pp. 445-507, 1992.

\bibitem{sauer_iterative_1987}
K. D. Sauer, J. P. Allebach, ``Iterative reconstruction of bandlimited images from nonuniformly spaced samples,''
\emph{IEEE Trans. Circuits and Syst.}, vol. 34, no. 12, pp. 1497-1506, 1987.

\bibitem{marvasti_nonuniform_2001}
F. Marvasti, \emph{Nonuniform Sampling: Theory and Practice}, Springer, 2001.

\bibitem{marvasti_recovery_1991}
F. Marvasti, M. Analoui, and M. Gamshadzahi, ``Recovery of signals from nonuniform samples using iterative methods,''
\emph{IEEE Trans. Signal Process.}, vol. 39, no. 4, pp. 872-878, 1991.

\bibitem{grochenig_reconstruction_1992}
K. Gr\"{o}chenig, ``Reconstruction algorithms in irregular sampling,''
\emph{Mathematics of Computation}, vol. 59, no. 199, pp. 181-194, 1992.

\bibitem{IntelData}
\url{http://www.cs.cmu.edu/~guestrin/Research/Data/}.

\end{thebibliography}
\end{document}